\input harvmac
\def\ev#1{\langle#1\rangle}
\input amssym
\input epsf

\def\unit{\relax{\rm 1\kern-.26em I}}
\def\nada{\relax{\rm 0\kern-.30em l}}



\def\det{{\rm det}}

\noblackbox
\def\IL{\relax{\rm I\kern-.18em L}}
\def\IH{\relax{\rm I\kern-.18em H}}
\def\IR{\relax{\rm I\kern-.18em R}}
\def\IC{\relax\hbox{$\inbar\kern-.3em{\rm C}$}}
\def\IZ{\relax\ifmmode\mathchoice
{\hbox{\cmss Z\kern-.4em Z}}{\hbox{\cmss Z\kern-.4em Z}}
{\lower.9pt\hbox{\cmsss Z\kern-.4em Z}} {\lower1.2pt\hbox{\cmsss
Z\kern-.4em Z}}\else{\cmss Z\kern-.4em Z}\fi}
\def\CM {{\cal M}}
\def\CN {{\cal N}}

\def\CL {{\cal L}}

\def\CO {{\cal O}}

\def\CM {{\cal M}}
\def\CN {{\cal N}}

\def\CO {{\cal O}}

\def\det{{\rm det}}
\def\Tr{{\rm Tr}}

\font\manual=manfnt \def\dbend{\lower3.5pt\hbox{\manual\char127}}

\def\IZ{\relax\ifmmode\mathchoice
{\hbox{\cmss Z\kern-.4em Z}}{\hbox{\cmss Z\kern-.4em Z}}
{\lower.9pt\hbox{\cmsss Z\kern-.4em Z}} {\lower1.2pt\hbox{\cmsss
Z\kern-.4em Z}}\else{\cmss Z\kern-.4em Z}\fi}
\def\half {{1\over 2}}

\def\rt2{\sqrt{2}}
\def\irt2{{1\over\sqrt{2}}}

\def\hat{\widehat}
\def\slashchar#1{\setbox0=\hbox{$#1$}           
   \dimen0=\wd0                                 
   \setbox1=\hbox{/} \dimen1=\wd1               
   \ifdim\dimen0>\dimen1                        
      \rlap{\hbox to \dimen0{\hfil/\hfil}}      
      #1                                        
   \else                                        
      \rlap{\hbox to \dimen1{\hfil$#1$\hfil}}   
      /                                         
   \fi}

\def\foursqr#1#2{{\vcenter{\vbox{
    \hrule height.#2pt
    \hbox{\vrule width.#2pt height#1pt \kern#1pt
    \vrule width.#2pt}
    \hrule height.#2pt
    \hrule height.#2pt
    \hbox{\vrule width.#2pt height#1pt \kern#1pt
    \vrule width.#2pt}
    \hrule height.#2pt
        \hrule height.#2pt
    \hbox{\vrule width.#2pt height#1pt \kern#1pt
    \vrule width.#2pt}
    \hrule height.#2pt
        \hrule height.#2pt
    \hbox{\vrule width.#2pt height#1pt \kern#1pt
    \vrule width.#2pt}
    \hrule height.#2pt}}}}
\def\psqr#1#2{{\vcenter{\vbox{\hrule height.#2pt
    \hbox{\vrule width.#2pt height#1pt \kern#1pt
    \vrule width.#2pt}
    \hrule height.#2pt \hrule height.#2pt
    \hbox{\vrule width.#2pt height#1pt \kern#1pt
    \vrule width.#2pt}
    \hrule height.#2pt}}}}
\def\sqr#1#2{{\vcenter{\vbox{\hrule height.#2pt
    \hbox{\vrule width.#2pt height#1pt \kern#1pt
    \vrule width.#2pt}
    \hrule height.#2pt}}}}
\def\square{\mathchoice\sqr65\sqr65\sqr{2.1}3\sqr{1.5}3}
\def\doub{\mathchoice\psqr65\psqr65\psqr{2.1}3\psqr{1.5}3}

\lref\HananyNA{
  A.~Hanany and Y.~Oz,
  ``On the quantum moduli space of vacua of N=2 supersymmetric SU(N(c)) gauge
  theories,''
  Nucl.\ Phys.\ B {\bf 452}, 283 (1995)
  [arXiv:hep-th/9505075].
}

\lref\ArgyresCB{
  P.~C.~Argyres and M.~Edalati,
  ``Generalized Konishi anomaly, Seiberg duality and singular effective
  superpotentials,''
  arXiv:hep-th/0511272.
}

\lref\tyui{A.~Hanany and Y.~Oz, ``On the quantum moduli space of
vacua of N=2 supersymmetric SU(N(c)) gauge theories,'' Nucl.\
Phys.\ B {\bf 452}, 283 (1995) [arXiv:hep-th/9505075].
}
\lref\tyuii{P.~C.~Argyres, M.~R.~Plesser and A.~D.~Shapere, ``The
Coulomb phase of N=2 supersymmetric QCD,'' Phys.\ Rev.\ Lett.\
{\bf 75}, 1699 (1995) [arXiv:hep-th/9505100].
}
\lref\tyuiii{J.~A.~Minahan and D.~Nemeschansky, ``Hyperelliptic
curves for supersymmetric Yang-Mills,'' Nucl.\ Phys.\ B {\bf 464},
3 (1996) [arXiv:hep-th/9507032].
}
\lref\tyuiv{I.~M.~Krichever and D.~H.~Phong, ``On the integrable
geometry of soliton equations and N = 2  supersymmetric gauge
theories,'' J.\ Diff.\ Geom.\  {\bf 45}, 349 (1997)
[arXiv:hep-th/9604199].
}
\lref\tyuv{E.~D'Hoker, I.~M.~Krichever and D.~H.~Phong, ``The
effective prepotential of N = 2 supersymmetric SU(N(c)) gauge
theories,'' Nucl.\ Phys.\ B {\bf 489}, 179 (1997)
[arXiv:hep-th/9609041].
}

\lref\ISS{
  K.~A.~Intriligator, N.~Seiberg and S.~H.~Shenker,
  ``Proposal for a simple model of dynamical SUSY breaking,''
  Phys.\ Lett.\ B {\bf 342}, 152 (1995)
  [arXiv:hep-ph/9410203].
}
\lref\DineYW{
  M.~Dine and A.~E.~Nelson,
  ``Dynamical supersymmetry breaking at low-energies,''
  Phys.\ Rev.\ D {\bf 48}, 1277 (1993)
  [arXiv:hep-ph/9303230].
}
\lref\AffleckUZ{
  I.~Affleck, M.~Dine and N.~Seiberg,
  ``Calculable Nonperturbative Supersymmetry Breaking,''
  Phys.\ Rev.\ Lett.\  {\bf 52}, 1677 (1984).
}
\lref\AffleckXZ{
  I.~Affleck, M.~Dine and N.~Seiberg,
  ``Dynamical Supersymmetry Breaking In Four-Dimensions And Its
  Phenomenological Implications,''
  Nucl.\ Phys.\ B {\bf 256}, 557 (1985).
}
\lref\GiudiceBP{
  G.~F.~Giudice and R.~Rattazzi,
  ``Theories with gauge-mediated supersymmetry breaking,''
  Phys.\ Rept.\  {\bf 322}, 419 (1999)
  [arXiv:hep-ph/9801271].
}
\lref\WittenKV{
  E.~Witten,
  ``Mass Hierarchies In Supersymmetric Theories,''
  Phys.\ Lett.\ B {\bf 105}, 267 (1981).
}
\lref\LutyVR{
  M.~A.~Luty and J.~Terning,
  ``Improved single sector supersymmetry breaking,''
  Phys.\ Rev.\ D {\bf 62}, 075006 (2000)
  [arXiv:hep-ph/9812290].
}

\lref\BaggerHH{
  J.~Bagger, E.~Poppitz and L.~Randall,
  ``The R axion from dynamical supersymmetry breaking,''
  Nucl.\ Phys.\ B {\bf 426}, 3 (1994)
  [arXiv:hep-ph/9405345].
}
\lref\WessCP{
  J.~Wess and J.~Bagger,
  ``Supersymmetry and supergravity,''
}
\lref\NSd{
  N.~Seiberg,
  ``Electric - magnetic duality in supersymmetric nonAbelian gauge theories,''
  Nucl.\ Phys.\ B {\bf 435}, 129 (1995)
  [arXiv:hep-th/9411149].
}
\lref\ArkaniHamedFQ{
  N.~Arkani-Hamed, M.~A.~Luty and J.~Terning,
  ``Composite quarks and leptons from dynamical supersymmetry breaking  without
  messengers,''
  Phys.\ Rev.\ D {\bf 58}, 015004 (1998)
  [arXiv:hep-ph/9712389].
}
\lref\WittenDF{
  E.~Witten,
  ``Constraints On Supersymmetry Breaking,''
  Nucl.\ Phys.\ B {\bf 202}, 253 (1982).
}

\lref\AffleckMK{
  I.~Affleck, M.~Dine and N.~Seiberg,
  ``Dynamical Supersymmetry Breaking In Supersymmetric QCD,''
  Nucl.\ Phys.\ B {\bf 241}, 493 (1984).
}

\lref\AffleckXZ{
  I.~Affleck, M.~Dine and N.~Seiberg,
  ``Dynamical Supersymmetry Breaking In Four-Dimensions And Its
  Phenomenological Implications,''
  Nucl.\ Phys.\ B {\bf 256}, 557 (1985).
}
\lref\CachazoJY{
  F.~Cachazo, K.~A.~Intriligator and C.~Vafa,
  ``A large N duality via a geometric transition,''
  Nucl.\ Phys.\ B {\bf 603}, 3 (2001)
  [arXiv:hep-th/0103067].
}
\lref\CachazoRY{
  F.~Cachazo, M.~R.~Douglas, N.~Seiberg and E.~Witten,
  ``Chiral rings and anomalies in supersymmetric gauge theory,''
  JHEP {\bf 0212}, 071 (2002)
  [arXiv:hep-th/0211170].
}
\lref\BanksRU{
  T.~Banks and M.~Johnson,
  ``Regulating eternal inflation,''
  arXiv:hep-th/0512141.
}

\lref\DijkgraafDH{
  R.~Dijkgraaf and C.~Vafa,
  ``A perturbative window into non-perturbative physics,''
  arXiv:hep-th/0208048.
}

\lref\SWii{
  N.~Seiberg and E.~Witten,
  ``Monopoles, duality and chiral symmetry breaking in N=2 supersymmetric
  QCD,''
  Nucl.\ Phys.\ B {\bf 431}, 484 (1994)
  [arXiv:hep-th/9408099].
}
\lref\GGRR{
  G.~F.~Giudice and R.~Rattazzi,
  ``Extracting supersymmetry-breaking effects from wave-function
  renormalization,''
  Nucl.\ Phys.\ B {\bf 511}, 25 (1998)
  [arXiv:hep-ph/9706540].
}
\lref\AFGJ{D.~Anselmi, D.~Z.~Freedman, M.~T.~Grisaru and A.~A.~Johansen,
``Nonperturbative formulas for central functions of supersymmetric gauge
theories,''
Nucl.\ Phys.\ B {\bf 526}, 543 (1998)
[arXiv:hep-th/9708042].
}
\lref\BaggerHH{
  J.~Bagger, E.~Poppitz and L.~Randall,
  ``The R axion from dynamical supersymmetry breaking,''
  Nucl.\ Phys.\ B {\bf 426}, 3 (1994)
  [arXiv:hep-ph/9405345].
}
\lref\ColemanPY{
  S.~R.~Coleman,
  ``The Fate Of The False Vacuum. 1. Semiclassical Theory,''
  Phys.\ Rev.\ D {\bf 15}, 2929 (1977)
  [Erratum-ibid.\ D {\bf 16}, 1248 (1977)].
}
\lref\LutyVR{
  M.~A.~Luty and J.~Terning,
  ``Improved single sector supersymmetry breaking,''
  Phys.\ Rev.\ D {\bf 62}, 075006 (2000)
  [arXiv:hep-ph/9812290].
}

\lref\BanksDF{
  T.~Banks,
  ``Cosmological supersymmetry breaking and the power of the pentagon: A model
  of low energy particle physics,''
  arXiv:hep-ph/0510159.
}

\lref\DineAG{
  M.~Dine, A.~E.~Nelson, Y.~Nir and Y.~Shirman,
  ``New tools for low-energy dynamical supersymmetry breaking,''
  Phys.\ Rev.\ D {\bf 53}, 2658 (1996)
  [arXiv:hep-ph/9507378].
}

\lref\AffleckUZ{
  I.~Affleck, M.~Dine and N.~Seiberg,
  ``Calculable Nonperturbative Supersymmetry Breaking,''
  Phys.\ Rev.\ Lett.\  {\bf 52}, 1677 (1984).
}
\lref\ITii{
  K.~A.~Intriligator and S.~D.~Thomas,
  ``Dual descriptions of supersymmetry breaking,''
  arXiv:hep-th/9608046.
}
\lref\IY{
  K.~I.~Izawa and T.~Yanagida,
  ``Dynamical Supersymmetry Breaking in Vector-like Gauge Theories,''
  Prog.\ Theor.\ Phys.\  {\bf 95}, 829 (1996)
  [arXiv:hep-th/9602180].
}
\lref\Chacko{
  Z.~Chacko, M.~A.~Luty and E.~Ponton,
  ``Calculable dynamical supersymmetry breaking on deformed moduli spaces,''
  JHEP {\bf 9812}, 016 (1998)
  [arXiv:hep-th/9810253].
}
\lref\IT{
  K.~Intriligator and S.~D.~Thomas,
  ``Dynamical Supersymmetry Breaking on Quantum Moduli Spaces,''
  Nucl.\ Phys.\ B {\bf 473}, 121 (1996)
  [arXiv:hep-th/9603158].
}
\lref\BarnesZN{
  E.~Barnes, K.~Intriligator, B.~Wecht and J.~Wright,
  ``N = 1 RG flows, product groups, and a-maximization,''
  Nucl.\ Phys.\ B {\bf 716}, 33 (2005)
  [arXiv:hep-th/0502049].
}
\lref\LeighSJ{
  R.~G.~Leigh, L.~Randall and R.~Rattazzi,
  ``Unity of supersymmetry breaking models,''
  Nucl.\ Phys.\ B {\bf 501}, 375 (1997)
  [arXiv:hep-ph/9704246].
}

\lref\CachazoJY{
  F.~Cachazo, K.~A.~Intriligator and C.~Vafa,
  ``A large N duality via a geometric transition,''
  Nucl.\ Phys.\ B {\bf 603}, 3 (2001)
  [arXiv:hep-th/0103067].
}

\lref\DuncanAI{
  M.~J.~Duncan and L.~G.~Jensen,
  ``Exact tunneling solutions in scalar field theory,''
  Phys.\ Lett.\ B {\bf 291}, 109 (1992).
}

\lref\IntriligatorFK{
  K.~A.~Intriligator and S.~Thomas,
  ``Dual descriptions of supersymmetry breaking,''
  arXiv:hep-th/9608046.
}

\lref\SWi{
  N.~Seiberg and E.~Witten,
  ``Electric - magnetic duality, monopole condensation, and confinement in N=2
  supersymmetric Yang-Mills theory,''
  Nucl.\ Phys.\ B {\bf 426}, 19 (1994)
  [Erratum-ibid.\ B {\bf 430}, 485 (1994)]
  [arXiv:hep-th/9407087].
}

\lref\DKi{
D.~Kutasov,
``A Comment on duality in N=1 supersymmetric nonAbelian gauge
theories,''
Phys.\ Lett.\ B {\bf 351}, 230 (1995)
[arXiv:hep-th/9503086].
}

\lref\DKAS{
D.~Kutasov and A.~Schwimmer,
``On duality in supersymmetric Yang-Mills theory,''
Phys.\ Lett.\ B {\bf 354}, 315 (1995)
[arXiv:hep-th/9505004].
}

\lref\DKNSAS{D.~Kutasov, A.~Schwimmer and N.~Seiberg,
``Chiral Rings, Singularity Theory and Electric-Magnetic Duality,''
Nucl.\ Phys.\ B {\bf 459}, 455 (1996)
[arXiv:hep-th/9510222].
}

\lref\BrodieVX{
  J.~H.~Brodie,
  ``Duality in supersymmetric SU(N/c) gauge theory with two adjoint chiral
  superfields,''
  Nucl.\ Phys.\ B {\bf 478}, 123 (1996)
  [arXiv:hep-th/9605232].
}

\lref\PouliotSK{
  P.~Pouliot and M.~J.~Strassler,
  ``A Chiral $SU(N)$ Gauge Theory and its Non-Chiral $Spin(8)$ Dual,''
  Phys.\ Lett.\ B {\bf 370}, 76 (1996)
  [arXiv:hep-th/9510228].
}
\lref\IntriligatorNE{
  K.~A.~Intriligator and P.~Pouliot,
  ``Exact superpotentials, quantum vacua and duality in supersymmetric SP(N(c))
  gauge theories,''
  Phys.\ Lett.\ B {\bf 353}, 471 (1995)
  [arXiv:hep-th/9505006].
 }
\lref\IntriligatorID{
  K.~A.~Intriligator and N.~Seiberg,
  ``Duality, monopoles, dyons, confinement and oblique confinement in
  supersymmetric SO(N(c)) gauge theories,''
  Nucl.\ Phys.\ B {\bf 444}, 125 (1995)
  [arXiv:hep-th/9503179].
}
\lref\ISrev{
  K.~A.~Intriligator and N.~Seiberg,
  ``Lectures on supersymmetric gauge theories and electric-magnetic  duality,''
  Nucl.\ Phys.\ Proc.\ Suppl.\  {\bf 45BC}, 1 (1996)
  [arXiv:hep-th/9509066].
}
\lref\NelsonNF{
  A.~E.~Nelson and N.~Seiberg,
  ``R symmetry breaking versus supersymmetry breaking,''
  Nucl.\ Phys.\ B {\bf 416}, 46 (1994)
  [arXiv:hep-ph/9309299].
}
\lref\DineAG{
  M.~Dine, A.~E.~Nelson, Y.~Nir and Y.~Shirman,
  ``New tools for low-energy dynamical supersymmetry breaking,''
  Phys.\ Rev.\ D {\bf 53}, 2658 (1996)
  [arXiv:hep-ph/9507378].
}
\lref\SeibergBZ{
  N.~Seiberg,
  ``Exact results on the space of vacua of four-dimensional SUSY gauge
  theories,''
  Phys.\ Rev.\ D {\bf 49}, 6857 (1994)
  [arXiv:hep-th/9402044].
}

\lref\SeibergRS{
  N.~Seiberg and E.~Witten,
  ``Electric - magnetic duality, monopole condensation, and confinement in N=2
  supersymmetric Yang-Mills theory,''
  Nucl.\ Phys.\ B {\bf 426}, 19 (1994)
  [Erratum-ibid.\ B {\bf 430}, 485 (1994)]
  [arXiv:hep-th/9407087].
}

\lref\IntriligatorFK{
  K.~A.~Intriligator and S.~D.~Thomas,
  ``Dual descriptions of supersymmetry breaking,''
  arXiv:hep-th/9608046.
}
\lref\LeighSJ{
  R.~G.~Leigh, L.~Randall and R.~Rattazzi,
  ``Unity of supersymmetry breaking models,''
  Nucl.\ Phys.\ B {\bf 501}, 375 (1997)
  [arXiv:hep-ph/9704246].
}

\newbox\tmpbox\setbox\tmpbox\hbox{\abstractfont }
\Title{\vbox{\baselineskip12pt \hbox{UCSD-PTH-06-03}}}
{\vbox{\centerline{Dynamical SUSY Breaking in Meta-Stable Vacua}}}
\smallskip
\centerline{Kenneth Intriligator$^{1,2}$, Nathan Seiberg$^2$ and
David Shih$^3$}
\smallskip
\bigskip
\centerline{$^1${\it Department of Physics, University of
California, San Diego, La Jolla, CA 92093 USA}}
\medskip
\centerline{$^2${\it School of Natural Sciences, Institute for
Advanced Study, Princeton, NJ 08540 USA}}
\medskip
\centerline{$^3${\it Department of Physics, Princeton University,
Princeton, NJ 08544 USA}}
\bigskip
\vskip 1cm

\noindent Dynamical supersymmetry breaking in a long-lived
meta-stable vacuum is a phenomenologically viable possibility.
This relatively unexplored avenue leads to many new models of
dynamical supersymmetry breaking. Here, we present a surprisingly
simple class of models with meta-stable dynamical supersymmetry
breaking: $\CN =1$ supersymmetric QCD, with massive flavors.
Though these theories are strongly coupled, we definitively
demonstrate the existence of meta-stable vacua by using the
free-magnetic dual. Model building challenges, such as large
flavor symmetries and the absence of an R-symmetry, are easily
accommodated in these theories. Their simplicity also suggests
that broken supersymmetry is generic in supersymmetric field
theory and in the landscape of string vacua.

\bigskip

\Date{February 2006}

\newsec{Introduction}

\subsec{General Remarks}

At first glance, dynamical supersymmetry breaking appears to be a
rather non-generic phenomenon in supersymmetric gauge theory. The
non-zero Witten index of $\CN =1$ Yang-Mills theory immediately
implies that any $\CN =1$ supersymmetric gauge theory with
massive, vector-like matter has supersymmetric vacua \WittenDF. So
theories with no supersymmetric vacua must either be chiral, as in
the original examples of \refs{\AffleckUZ, \AffleckXZ}, or if they
are non-chiral, they must have massless matter, as in the examples
of \refs{\IT, \IY}. The known theories that satisfy these
requirements and dynamically break supersymmetry look rather
complicated, and applications to realistic model building only
compounds the complications. The result has been a literature of
rather baroque models of dynamical supersymmetry breaking and
mediation.  For reviews and references, see e.g.\ \GiudiceBP.

We point out that new model building avenues are opened up by
abandoning the prejudice that models of dynamical supersymmetry
breaking must have {\it no} supersymmetric vacua. This prejudice
is unnecessary, because it is a phenomenologically viable
possibility that we happen to reside in a very long lived, false
vacuum, and that there is a supersymmetric vacuum elsewhere in
field space.  Meta-stable supersymmetry breaking vacua have been
encountered before in the literature of models of supersymmetry
breaking and mediation; some examples are \refs{\DineAG\LutyVR -
\BanksDF}.   Indeed, even if the supersymmetry breaking sector has
no supersymmetric vacua, there is a danger that the mediation
sector will introduce supersymmetric vacua elsewhere. Such
encounters of meta-stable supersymmetry breaking are generally
accompanied with a (justified)
apology for the aesthetic defect and, in favorable cases, it is
shown that the lifetime can nevertheless be longer than the age of
the Universe.

The novelty here is that we accept meta-stable vacua from the
outset, even in the supersymmetry breaking sector. This approach
leads us immediately to many new and much simpler models of
supersymmetry breaking. Classic constraints, needed for having
{\it no} supersymmetric vacua, no longer constrain models of
meta-stable supersymmetry breaking. For instance, theories with
non-zero Witten index and/or with no conserved $U(1)_R$ symmetry
\refs{\AffleckXZ, \NelsonNF}\ can nevertheless have meta-stable
supersymmetry breaking vacua. A condition for supersymmetry
breaking that does still apply in the meta-stable context is the
need for a massless fermion to play the role of the Goldstino. But
even this condition can be subtle: the massless fermion can be
present in the low-energy macroscopic theory, even if it is not
obvious in the original, ultraviolet, microscopic theory.   This
happens in our examples.

Phenomenologically, we would like the lifetime of our meta-stable
state to be longer than the age of the Universe. Moreover, the
notion of meta-stable states is meaningful only when they are
parametrically long lived.  It is therefore important for us to
have a dimensionless parameter, $\epsilon$, whose parametric
smallness guarantees the longevity of the meta-stable state.  In our examples,
$\epsilon$ is given by a ratio of a mass and a dynamical scale,
\eqn\epsdef{\epsilon \equiv {\mu \over \Lambda _m}\sim \sqrt{{m\over \Lambda}},}
where the masses and scales will be explained shortly.
What happens to the meta-stable state as $\epsilon \to 0$ depends on
what we hold fixed.  In some examples, we should hold the dynamical scale $\Lambda$ fixed,
then as $\epsilon \to 0$, the meta-stable state becomes supersymmetric. In
other examples, we should hold the mass scale $\mu$
fixed, and then supersymmetry is
broken.

Most of the analysis of supersymmetry dynamics in the past has
been concerned with BPS / chiral / holomorphic quantities, which
are protected in some way by supersymmetry. Since we are
interested in supersymmetry breaking, we have to go outside this
domain, and our answers depend on non-chiral information which in
general cannot be computed.  In the past, calculable models of
dynamical supersymmetry breaking were based on the fact that the
vacuum ended up being at large fields, where the K\"ahler
potential is approximately classical for the fields of the
microscopic theory \AffleckXZ. In this paper we study vacua at
small field expectation values, where the K\"ahler potential is
complicated. Here our small parameter $\epsilon$ will
be useful. Taking $\epsilon \rightarrow 0$, holding fixed the dynamical
scale $\Lambda$, supersymmetry is
unbroken and we know the spectrum of the IR theory.  When this
theory is IR free, the K\"ahler metric of the light modes is
smooth and it can be parameterized by a small number of real
coefficients of order one.  Even though we do not know how to
compute these coefficients, we will be able to express a lot of
information (the ground state energy, the spectrum of light
particles, the effective potential, etc.) in terms of them. This
approach has already been used in \Chacko, to analyze the
supersymmetry breaking model of \refs{\IT, \IY}\ in the strong
coupling region.

\subsec{Our main example}

Our main example of meta-stable dynamical supersymmetry breaking
in this paper is surprisingly simple: $\CN =1$ supersymmetric
$SU(N_c)$ QCD, with $N_f$ massive fundamental flavors. In order to
have control over the theory in the IR, we take $N_f$ in the free
magnetic range \refs{\SeibergBZ \NSd - \ISrev}, $N_c+1\leq N_f<
{3\over 2}N_c$. We will show that, in addition to the expected
supersymmetric vacua of a theory with massive vector-like matter,
there are {\it long-lived non-supersymmetric vacua}. Our analysis
is reliable in a particular limit,
 \eqn\epsdef{
 |\epsilon |\sim \sqrt{\left|{m\over\Lambda}\right|} \ll 1.
 }
where $m$ is the typical scale of the quark masses and $\Lambda$
is the strong-coupling scale of the theory. Using the
free-magnetic dual description of the theory in the infrared, we
determine properties of the strongly coupled gauge theory outside
of the usual realm of holomorphic quantities and supersymmetric
vacua. The simplicity of these models leads us to suspect that
meta-stable vacua with broken supersymmetry are generic.

In the infrared description of the theory, supersymmetry is spontaneously broken
at tree-level by what we refer to as the ``rank-condition"
mechanism of supersymmetry breaking. Consider a theory of chiral
superfields $\Phi_{ij}$, $\varphi ^i_c$, and $\tilde \varphi
^{ic}$, with $i=1\dots N_f$, and $c=1\dots N$, with $N<N_f$, and
tree-level superpotential
 \eqn\wmqq{W=h\Tr\, \varphi \Phi \tilde \varphi -h\mu ^2\Tr\, \Phi.}
The F-terms of $\Phi$, $F_{\Phi _{ij}}\sim \tilde \varphi ^{jc}\varphi _c\,{}^i-h\mu ^2\delta ^{ij}$,
cannot all vanish, because $\delta ^{ij}$ has rank $N_f$ but  $\tilde \varphi ^{jc}\varphi _c\,{}^i$
only has rank $N<N_f$.  Supersymmetry is thus spontaneously broken. For $SU(N_c)$ SQCD with
$N_c+1\le N_f<{3\over2}N_c$, \wmqq\ arises as the infrared
free, low-energy effective theory of the magnetic dual \NSd, with
$N=N_f-N_c$ and $\mu \sim \sqrt{m\Lambda}$.

At tree-level in the macroscopic theory \wmqq, there is a moduli
space of degenerate, non-supersymmetric vacua, labelled by
arbitrary expectation values of some classically massless fields,
which are some components of the fields in \wmqq. Some of these
fields are Goldstone bosons of broken global symmetries, and
remain as exactly massless moduli of the vacua. (The moduli space,
being of the form $G/H$, is always compact.) There are also
classically massless ``pseudo-moduli"; these get a potential from
perturbative quantum corrections in the effective theory \wmqq.
The leading perturbative contribution to the potential for the
pseudo-moduli can be computed using the one-loop correction to the
vacuum energy,
 \eqn\CWgen{
 V^{(1)}_{eff} = {1\over
 64\pi^2}{\rm STr}\,\CM^4\log{\CM^2\over\Lambda ^2}\equiv
 {1\over 64\pi ^2}\left( \Tr \, m_B^4 \log {m_B^2\over \Lambda ^2}
 -\Tr \, m_F^4 \log {m_F^2\over \Lambda ^2}\right),}
where $m_B^2$ and $m_F^2$ are the tree-level boson and fermion
masses, as a function of the expectation values of the
pseudo-moduli.\foot{The ultraviolet cutoff $\Lambda$  in \CWgen\
can be absorbed into the renormalization of the coupling constants
appearing in the tree-level vacuum energy $V_0$. In particular,
${\rm STr}\,\CM^4$ is independent of the pseudo-moduli.} Using
\CWgen, we find non-supersymmetric vacua, stabilized by a
potential barrier which to leading order scales like $|\mu ^2|$.
In terms of the parameter $\mu ^2$ appearing in \wmqq, this
effective potential is thus not real analytic. That  is why this
potential, computed in the low-energy macroscopic theory, is
robust upon including effects from the underlying microscopic
theory.  We will discuss this in more detail below.

The $N_c$ supersymmetric vacua expected from the Witten index of
$SU(N_c)$ SQCD with massive matter can also be seen in the
low-energy macroscopic theory of the free magnetic dual. Giving
the fields $\Phi$ in \wmqq\ expectation values, gaugino
condensation in the $SU(N)$ magnetic dual contributes to the
superpotential and leads to the expected $N_c$ supersymmetric
vacua. This is an interesting example of non-perturbative {\it
restoration} of supersymmetry in a theory which breaks
supersymmetry at tree-level.

\subsec{Outline}

As we have summarized, our microscopic UV theory is $SU(N_c)$
SQCD, and we analyze its supersymmetry-breaking dynamics using the
macroscopic, IR-free dual. In the body of the paper, we will
follow a bottom-up presentation, starting in the IR, and then
working up to the UV. The advantage of this bottom up approach is
that, as we shall discuss, the important physics of supersymmetry
breaking all happens in the infrared theory. Effects from the
underlying, microscopic theory do not significantly affect the
conclusions. These considerations apply more broadly than to the
particular models that we analyze here.

In section 2, we discuss the rank-condition supersymmetry breaking
in the macroscopic, low-energy theory \wmqq, taking $SU(N)$ to be
a global, rather than gauge symmetry.  We compute the leading
effect from the one-loop potential \CWgen. These theories have
absolutely stable, non-supersymmetric vacua.  In section 3, we
gauge the $SU(N)$ group, taking $N_f>3N$ (which becomes $N_f<
{3\over2}N_c$ in the electric theory, after using $N=N_f-N_c$) so
the theory is infrared free.  The $SU(N)$ gauge group is
completely Higgsed in the non-supersymmetric vacua, and the
leading quantum effective potential is essentially the same as
that found in section 2.  The $SU(N)$ gauge fields do not much
affect the non-supersymmetric vacua, but they do have an important
effect elsewhere in field space, where they lead to
non-perturbative restoration of supersymmetry.  So the
non-supersymmetric vacua are only meta-stable, once $SU(N)$ is
gauged.

In section 4, we provide a short, general discussion on why it is
valid to take a bottom up approach, analyzing supersymmetry
breaking and the vacuum in the low-energy, macroscopic effective
theory.  It is argued in general that effects from the underlying
microscopic theory, whatever they happen to be, do not
significantly affect the conclusions.

In section 5, we connect the macroscopic effective field theories,
studied in the previous sections, with a microscopic description
in terms of $SU(N_c)$ SQCD with $N_f$ fundamental flavors.  The
fields $\Phi_{ij}$ and $\varphi ^i$ and $\tilde \varphi ^j$ are
composite objects of the microscopic theory. As discussed in
section 4, strong quantum effects of the underlying microscopic
theory do not alter our conclusions about the meta-stable
supersymmetry breaking vacuum.

In section 6, we discuss analogous models of meta-stable
supersymmetry breaking, based on $SO(N)$ (or more precisely,
$Spin(N)$) and $Sp(N)$ groups with fundamental matter. For the
case of $Spin(N)$, we argue that the meta-stable
non-supersymmetric vacua and the supersymmetric vacua are in
different phases: one is confining, and the other is oblique
confining.

In section 7, we show that our meta-stable vacua can be made
parametrically long lived. This makes them well defined and
phenomenologically interesting.  Finally, in section 8, we make
some preliminary comments about applications to model building.

In appendix A, we review some basic aspects of F-term
supersymmetry breaking. In appendix B, we provide some technical
details of the computation of the one-loop effective potential in
section 2. In appendix C, we present supersymmetric gauge
theories, based $SU(N)$ supersymmetric gauge theory with adjoint
matter, which have landscapes of supersymmetry breaking vacua.
Such gauge theories can naturally arise in string theory.   In
appendix D, we suggest testing for meta-stable non-supersymmetric
vacua in the context of $\CN =2$ supersymmetry, with small
explicit breaking to $\CN =1$, using the exactly known $\CN =2$
K\"ahler potential of \SeibergRS\ and following works.  For the
particular case of $SU(2)$ with no matter, we observe that there
is no meta-stable, non-supersymmetric vacuum.

\newsec{The Macroscopic Model: Part I}

In this section we discuss our macroscopic theory \wmqq\ without
the gauge interactions.  This is a Wess-Zumino model with global
symmetry group
 \eqn\globalsymgroup{ SU(N)\times SU(N_f)^2\times
 U(1)_B\times U(1)'\times U(1)_R }
(later we will identify $N=N_f-N_c$), with $N_f>N$ and the
following matter content
 \eqn\globalsym{
 \matrix{ & SU(N) & SU(N_f) & SU(N_f) & U(1)_B & U(1)' & U(1)_R \cr
          & \cr
         \Phi & 1 & \square  & \overline{\square} &  0 & -2 &  2\cr
         & \cr
         \varphi & \square & \overline{\square} & 1 & 1 & 1 & 0 \cr
         & \cr
         \tilde \varphi & \overline{\square} & 1 & \square & -1 & 1 & 0\cr
         }
}
We will take the canonical K\"ahler potential,
\eqn\Ksunge{
K =\Tr\,\varphi^\dagger \varphi+\Tr\,\tilde \varphi^\dagger \tilde
\varphi + \Tr\,\Phi^\dagger \Phi
}
and tree-level superpotential
\eqn\Wsunge{
W = h\Tr\, \varphi \Phi \tilde \varphi - h\mu^2\Tr\,\Phi .
 }
The first term in \Wsunge\ is the most general $W_{tree}$
consistent with the global symmetries \globalsymgroup. The
second term in \Wsunge\ breaks the global symmetry to
$SU(N)\times SU(N_f)\times U(1)_B\times U(1)_R$, where the
unbroken $SU(N_f)$ is the diagonal subgroup of the original
$SU(N_f)^2$.

Since $N_f>N$, the F-terms cannot be simultaneously set to zero,
and so supersymmetry is spontaneously broken by the rank
condition, as described in the introduction. The scalar potential
is minimized, with
 \eqn\Vminii{
V_{min} = (N_f-N)\left| h^2\mu^4\right| ,
 }
along a classical moduli space of vacua which,  up to global
symmetries, is given by
 \eqn\Vmin{
 \Phi =\pmatrix{0&0 \cr 0& \Phi_0}, \qquad \varphi=\pmatrix{\varphi_0\cr
0},\qquad \tilde \varphi ^T=\pmatrix{\tilde \varphi_0\cr 0},\qquad
{\rm with}\,\,\,\,
\tilde\varphi_0\varphi_0 = \mu^2\unit_{N}.
 }
Here $\Phi_0$ is an arbitrary $(N_f-N)\times(N_f-N)$ matrix, and
$\varphi_0$ and $\tilde \varphi_0$ are $N\times N$ matrices (the
zero entries in \Vmin\ are matrices).

The vacua of maximal unbroken global symmetry are (up to unbroken
flavor rotations)
 \eqn\maximal{
 \Phi_0=0,\qquad \varphi_0=\tilde \varphi_0=\mu\unit_{N},
 }
This preserves an unbroken $SU(N)_D\times SU(N_f-N)\times
U(1)_{B'}\times U(1)_R$, as well as a discrete charge conjugation symmetry that exchanges $\varphi$ and $\tilde \varphi$.

 We now examine the one-loop effective
potential of the classical pseudo-flat directions around the vacua \maximal.  To simplify
the presentation, we will expand around \maximal\ and show that
the classical pseudo-moduli there get positive mass-squared.

To see what the light fields are,we expand around \maximal\ using
the parametrization
\eqn\fluctsuni{
\Phi =\pmatrix{\delta Y& \delta Z^T\cr \delta\tilde Z& \delta
\hat\Phi}, \qquad
\varphi=\pmatrix{\mu+{1\over\sqrt{2}}(\delta\chi_+ +
\delta\chi_-)\cr {1\over \sqrt{2}}(\delta\rho _++\delta \rho
_-)},\qquad \tilde \varphi ^T=\pmatrix{
\mu+{1\over\sqrt{2}}(\delta\chi_+-\delta\chi_-)\cr {1\over
\sqrt{2}}(\delta\rho _+-\delta \rho _-)}
 }
(Here $\delta Y$ and $\delta \chi _\pm$ are $N\times N$ matrices,
and $\delta Z$, $\delta \tilde Z$, and $\delta \rho _\pm$ are
$(N_f-N)\times N$ matrices.) The potential from \Wsunge\ gives
most of the fields tree-level masses $\sim |h \mu|$.   There are
also massless scalars, some of which are Goldstone bosons of the
broken global symmetries:
 \eqn\goldstone{{\mu ^*\over |\mu|}\delta \chi _- - h.c., \qquad {\rm Re}\left({\mu^*\over |\mu|}
\delta \rho _+ \right), \qquad {\rm Im}\left({\mu ^*\over
|\mu|}\delta \rho _-\right).} The first is in
$SU(N)\times SU(N)_F\times U(1)_B/SU(N)_D$, and the
latter two are in $SU(N_f)/SU(N)_F\times SU(N_f-N)\times U(1)_{B'}$, where
$SU(N)_F\subset SU(N_f)$.

The other classically massless scalars are fluctuations of the
classical pseudo-flat directions,
 \eqn\hatchidef{
 \delta\hat\Phi\,\,\,\,{\rm and}\,\,\,\, \delta\hat\chi\equiv {\mu ^*\over |\mu|}\delta \chi _- + h.c.
 }
These pseudo-moduli acquire masses, starting at one-loop,
from their couplings to the massive fields. The effective theory
for the  pseudo-moduli has the form
 \eqn\eff{ \CL_{eff} = \Tr\,\partial(\delta
 \hat\Phi)^\dagger\partial(\delta\hat\Phi) +
 {1\over2}\Tr\,(\partial(\delta\hat\chi))^2- V_{eff}^{(1)}(\delta
 \hat\Phi,\delta\hat\chi) + \dots }
where $\dots$ denotes higher order derivative interactions, as
well as terms coming from two or more loops of the massive fields.
The one-loop contribution to the effective potential dominates
over higher loops, because the coupling $h$ is (marginally)
irrelevant in the infrared.  The kinetic terms in \eff\ are
inherited from the tree-level kinetic terms from \Ksunge\ of the
full theory, so they are diagonal and canonical to leading order.

The one-loop effective potential for the pseudo-moduli can be
computed from the one-loop correction \CWgen\ to the vacuum
energy, in the background where the pseudo-moduli have expectation
values. Expanding to quadratic order around the vacua \maximal,
the effective potential for the pseudo-moduli must be of the form
\eqn\Vonegen{
 V_{eff}^{(1)}= \left| h^4\mu^2\right| \left({1\over2}a\,
 \Tr\,\delta\hat\chi^2 + b\, \Tr\,\delta \hat\Phi^\dagger \delta
 \hat\Phi\right) + \dots ,}
for some numerical coefficients $a$ and $b$.  Here we used the
global symmetries and the fact that only single traces appear in
\CWgen\foot{Equivalently, it is easily verified that only planar
diagrams contribute at one loop.} to determine the field
dependence in \Vonegen.  The factor of $|h ^4\mu ^2|$ follows from
dimensional analysis and the fact that the classical masses in
${\cal M}$ are all proportional to $h$.  Substituting the
classical masses into \CWgen, the result is
 \eqn\abans{ a =  {\log
 4-1\over 8\pi^2}(N_f-N),\qquad b={\log 4-1\over 8\pi^2}N\ . }
Some details of the calculation of $a$ and $b$ are given in
appendix B, where we also show how our macroscopic model is related to an
O'Raifeartaigh-like model of supersymmetry breaking.
In any event, the precise values of $a$ and $b$ are not too
important; what matters for us is that they are both positive. The
leading order effective potential for the pseudo-moduli is
 \eqn\Vonege{
 V^{(1)}_{eff} =
 {|h^4 \mu^2|(\log 4-1)\over 8\pi^2}\left(
 {1\over2}(N_f-N)
 \Tr\,\delta\hat\chi^2
 +
 N\,\Tr\,\delta\hat\Phi^\dagger \delta\hat\Phi\right) + \dots,
 }
so the vacua \maximal\ are indeed stable, without any tachyonic
directions.

The spectrum of the theory in the vacuum \maximal\
has a hierarchy of mass scales, dictated by the (marginally)
irrelevant coupling $h$.  Some fields have tree-level masses $\sim
|h \mu|$.  The pseudo-moduli have masses $\sim |h^2\mu|$ from
\Vonege. The Goldstone bosons of the broken global symmetries of
course remain exactly massless; in particular, no quantum
corrections could drive them tachyonic. There is also an exactly
massless Goldstino, because supersymmetry is broken.

\newsec{The Macroscopic Model: Part II  -- Dynamical SUSY Restoration}

We now gauge the $SU(N)$ symmetry of the
previous section.  We are interested in the case $N_f>3N$, where
the $SU(N)$ theory is IR free instead of asymptotically free. Thus
the theory has a scale $\Lambda_m$, above which it is strongly
coupled. (The subscript $m$ on $\Lambda _m$ is for
``macroscopic.") The running of the holomorphic gauge coupling of
$SU(N)$ is given by
 \eqn\irfreerun{e^{-8\pi ^2/g^2(E)+i\theta }=\left({E\over \Lambda
 _m}\right)^{N_f-3N}.}
So $g$ runs to zero in the infrared, and the theory there can be
analyzed perturbatively.
In the ultraviolet, we encounter a Landau pole at $E=|\Lambda_m|$;
thus, for energies $E\sim |\Lambda _m|$ and above, the $SU(N)$
theory is not well defined. A different description of the theory
is then needed.

Having gauged $SU(N)$, the scalar potential is now $V=V_F+V_D$, where
$V_F$ is the $F$-term potential discussed in the previous section,
and $V_D$ is the $D$-term potential
 \eqn\Vscgauge{\eqalign{
 V_D &=  {1\over2} g^2\sum_A(\Tr\,\varphi^\dagger T_A
 \varphi-\Tr\,\tilde \varphi T_A \tilde \varphi^\dagger )^2.\cr
 }}
The D-term potential \Vscgauge\ vanishes in the vacua \maximal, so
\maximal\ remains as a minimum of the tree-level potential. The
$SU(N)$ gauge symmetry is completely Higgsed in this vacuum.
Through the super-Higgs mechanism, the $SU(N)$ gauge fields
acquire mass $g\mu$, the erstwhile Goldstone bosons ${\rm Im}(\mu
^*\delta \chi _-/|\mu|)'$ are eaten (the prime denotes the
traceless part), and the erstwhile pseudo-moduli $\delta \hat\chi
' = {\rm Re}(\mu^*\delta\chi_-/|\mu|)'$ get a non-tachyonic,
tree-level mass $g\mu$ from \Vscgauge.\foot{We could have also
gauged $U(N)\cong SU(N)\times U(1)_B$ in \globalsym, giving
$U(1)_B$ gauge coupling $g'$. Then the $U(1)_B$ vector multiplet
gets a tree-level supersymmetric mass $\sim g' \mu$ in the vacuum
\maximal, by the super Higgs mechanism. In particular, its trace
part, $\Tr\, \delta \hat \chi$ gets a non-tachyonic mass $\sim
g'\mu$ at tree-level.} Thus, the fields $\delta\hat\Phi$ and
$\Tr\,\delta\hat\chi$ remain as classical pseudo-moduli.

We should
compute the leading quantum effective potential for these
pseudo-moduli, as in the previous section, to determine whether
the vacua \maximal\ are stabilized, or develop tachyonic
directions. Actually, no new calculation is needed: the effect of the added
$SU(N)$ gauge fields drops out in the leading order effective
potential for the pseudo-moduli. The reason is that the tree-level
spectrum of the massive $SU(N)$ vector supermultiplet is
supersymmetric, so its additional contributions to the supertrace
of \CWgen\ cancel.
To see this, note that the $SU(N)$ gauge fields do not directly
couple to the supersymmetry breaking: the D-terms \Vscgauge\
vanish on the pseudo-flat space, and the non-zero expectation
values of $\varphi$ and $\tilde \varphi$, which give the $SU(N)$
gauge fields their masses, do not couple directly to any non-zero
$F$ terms.

We conclude that the leading order effective potential \Vonege\
for the
pseudo-moduli is unaffected by the gauging of $SU(N)$.
The vacua are as in \maximal, with broken supersymmetry and no tachyonic
directions.

Though gauging the $SU(N)$ does not much affect the supersymmetry
breaking vacua \maximal, it does have an important effect
elsewhere in field space: it leads to supersymmetric vacua.  To
see this, consider giving $\Phi$ general, non-zero expectation
values.  By the superpotential \Wsunge, this gives the $SU(N)$
fundamental flavors,  $\varphi$ and $\tilde \varphi$, mass
$\ev{h\Phi}$.  Below the energy scale $\ev{h\Phi}$, we can
integrate out these massive flavors.  The low-energy theory is
then $SU(N)$ pure Yang-Mills, with holomorphic coupling given by
 \eqn\glowrun{e^{-8\pi ^2/g^2(E)+i\theta }=\left({\Lambda _L\over E}
 \right)^{3N}={h^{N_f}\det\Phi\over \Lambda _m^{N_f-3N}E^{3N}.}
 }
In the last equality, we matched the running coupling to that
above the energy scale  $\ev{h\Phi}$, as given in \irfreerun.  The
low-energy theory has superpotential
 \eqn\Wfull{
 W_{low} = N(h^{N_f}\Lambda _m^{-(N_f-3N)}\det\,\Phi)^{1/{N}} -h
 \mu^2\Tr\,\Phi,
 }
where the first term comes from  $SU(N)$ gaugino condensation, upon using
\glowrun\ to relate $\Lambda _L$ to $\Lambda _m$.  We
stress that the appearance of $\Lambda _m$ in \Wfull\ does not
signify that we are including any effects coming {}from physics at
or above the ultraviolet cutoff $\Lambda _m$.  Rather, it appears
because we have expressed the infrared free coupling $g$ as in
\irfreerun.

Extremizing the superpotential \Wfull, we find $N_f-N$
supersymmetric vacua at
 \eqn\susymin{
 \langle h \Phi\rangle = \Lambda _m\epsilon ^{2N/(N_f-N)}\unit
 _{N_f}=\mu {1\over \epsilon ^{(N_f-3N)/(N_f-N)}}\unit_{N_f},
 \qquad \hbox{where}\qquad \epsilon \equiv {\mu \over \Lambda _m}.}
Note that, for $|\epsilon | \ll 1$,
 \eqn\ineqs{\left| \mu \right| \ll \left| \ev{h\Phi}\right| \ll \left| \Lambda _m\right| .}
Because $\ev{h\Phi}$ is well below the Landau pole at $\Lambda
_m$, this analysis in the low-energy, macroscopic theory is
justified and reliable.  As we will discuss in section 7, $\left|
\mu \right| \ll \left| \ev{h\Phi}\right| $ also guarantees the
longevity of the meta-stable, non-supersymmetric vacua
\maximal.

We see here an amusing phenomenon: dynamical supersymmetry
{\it restoration}, in a theory that breaks supersymmetry at
tree-level.  
For $\Lambda_m\to \infty$ with $\mu$ fixed, the theory breaks
supersymmetry. For $\Lambda_m$ large but finite (corresponding to
small but nonzero $\epsilon$), a supersymmetric vacuum comes in
from infinity.  The relevant non-perturbative effect arises in an
IR free gauge theory, and it can be reliably
computed.

The existence of these supersymmetric vacua elsewhere in field
space implies that the non-supersymmetric vacua of the previous
section become only meta-stable upon gauging $SU(N)$.
The model with gauged $SU(N)$ therefore exhibits meta-stable
supersymmetry breaking. We shall realize it dynamically in section
5.

 We note that our conclusions are in complete accord with
the connection of \refs{\AffleckXZ,\NelsonNF} between the
existence of a $U(1)_R$ symmetry and broken supersymmetry. The
theory of the previous section has a conserved $U(1)_R$ symmetry,
and it has broken supersymmetry.   In the theory of this section,
there is no conserved $U(1)_R$ symmetry, because it is anomalous
under the gauged $SU(N)$; this breaking is explicit in \Wfull.
Correspondingly,  there are supersymmetric vacua. For $\ev{\Phi}$
near the origin, the $SU(N)$ gauge theory is IR free, so the
$U(1)_R$ symmetry returns as an accidental symmetry of the
infrared theory. So supersymmetry breaking in our meta-stable
vacuum near the origin is related to the accidental R-symmetry
there.

\newsec{Effects from the underlying microscopic theory}

The theory we discussed in the previous sections is IR free and
therefore it cannot be a complete theory.  It breaks down at the
UV scale $|\Lambda_m|$ where its gauge interactions become large.
(The coupling $h$ in \Wsunge\ also has a Landau pole; for simplicity
we discuss only a single scale $|\Lambda_m|$.)  In this section we
will examine whether our results above depend on the physics at
the scale $|\Lambda_m|$ which we do not have under control.  The
only dimensionful parameter of the low energy theory is $\mu$ and
therefore, we will assume
 \eqn\epsilonde{|\epsilon |= \left|{\mu \over \Lambda_m}\right| \ll
 1}
We
will argue that the inequality \epsilonde\ guarantees that our
calculations above give the dominant effect in the low energy
theory.

The first effect that we should worry about is loops of modes from
the high energy theory.  These can be summarized by correction
terms in the effective K\"ahler potential, which at quartic order
take the typical form
 \eqn\kcancorr{\delta K  = {c \over |\Lambda_m |^2}
 \Tr (\Phi ^\dagger \Phi )^2+\dots,}
with $c$ being a dimensionless number of order one. The standard
decoupling argument is based on the fact that such high dimension
operators are suppressed by inverse powers of $|\Lambda_m |$ and
therefore they do not affect the dynamics of the low energy
theory.

Let us explore in more detail this fact and its relation to the
one-loop computation of the effective potential described in
section 2. There, we calculated the effect of supersymmetry
breaking mass terms on the low energy effective potential of the
pseudo-flat directions.  In that computation we focused on the
light fields, whose mass is of order $\mu$ (for simplicity, we set
$h=1$), and we neglected the modes with mass of order $\Lambda_m$.
Can the effect of these modes, whose masses are also split by
supersymmetry breaking, change our conclusion about the effective
potential?

Our one-loop effective potential \Vonege\ is   proportional to
$|\mu ^2|$, and is thus not real analytic in the parameter $\mu ^2$
appearing in the superpotential.  This non-analyticity is because
the modes that we integrated out become massless as $\mu \to 0$,
so their contribution to the effective potential is singular
there.  On the other hand, corrections from heavier modes, whose
masses are of order $\Lambda_m$, are necessarily real analytic in
$\mu ^2$. In
particular, the leading correction from the microscopic theory to
the mass of the pseudo-modulus must have coefficient $|\mu^2
|^2/|\Lambda _m|^2= |\mu ^2 \epsilon ^2| \ll |\mu ^2|$. Such corrections are much smaller
than our result  from the low-energy macroscopic theory.  One way
to see that is to integrate out the massive modes for $\mu=0$ and
summarize the effect in a correction to the K\"ahler potential as
in \kcancorr.  Then we can use this corrected K\"ahler potential
with the tree level superpotential to find the effect on the
pseudo-flat directions.  These corrections are $\sim |\mu
^2\epsilon ^2|$, and are negligible.

This fact is significant.  Without knowing the details of the
microscopic theory, we cannot determine these loop effects
involving modes with mass $\sim \Lambda_m$. We cannot even
determine the sign of the dimensionless coefficients like $c$ in
\kcancorr, and therefore we cannot determine whether they bend the
pseudo-flat directions upward or downward. Fortunately, these
effects which we cannot compute are smaller than the one loop
effects in the low energy theory which we can compute.  The latter
have the effect of stabilizing our vacuum.

Of course, this discussion about the {\it irrelevance of
irrelevant operators} which are suppressed by powers of
$\Lambda_m$ is obvious and trivial. However, in equation \Wfull\
we took into account a nonperturbative effect which leads to a
superpotential which is suppressed by powers of $\Lambda_m$.  We
are immediately led to ask two questions.  First, how come this
nonrenormalizable interaction is reliably computed even though it
depends on $\Lambda_m$?  Second, given that we consider this
interaction, why is it justified to neglect other terms as in
\kcancorr\ which are also suppressed by powers of $\Lambda_m$?

Let us first address the first question.  As in \irfreerun,
$\Lambda _m$ appears as a way to parameterize the infrared free
gauge coupling $g$, at energy scales below $|\Lambda _m|$. This is
conceptually different from the appearance of $|\Lambda _m|$ in
\kcancorr, which has to do with effects from the microscopic
theory, above the Landau pole scale. The superpotential \Wfull\ is
generated by low energy effects and therefore it is correctly
computed in the low energy effective theory.   As a check, the
resulting expectation value of $\Phi$ \susymin\ is much smaller
than $\Lambda_m$ and therefore it is reliably calculated.

Let us now turn to the second question, of how we can neglect
higher order corrections to the K\"ahler potential while keeping
the superpotential \Wfull.  The leading contribution of such terms
comes {}from corrections in the K\"ahler potential \kcancorr\ of
the schematic form $|\Phi|^4/|\Lambda_m|^2$.  The leading effect
of such corrections in the scalar potential are, schematically,
 \eqn\leadpotc{\Delta_{K} V_{eff} \sim \left|{\mu^2 \Phi
 \over \Lambda_m}\right|^2\sim \left| \mu ^2\epsilon ^2\right|
 \left| \Phi\right|^2,}
which for $|\epsilon| \ll 1$ are negligible corrections to the term
\Vonege\ that we computed above. Higher order corrections to the
K\"ahler potential are suppressed by even higher powers of
$\Phi\over \Lambda_m$, and are clearly negligible for $|\Phi | \ll
|\Lambda_m|$.  The correction \leadpotc\ should be compared with
the correction to the tree level potential from the superpotential
\Wfull, which is of the form
 \eqn\leadpotp{\Delta_{W} V_{eff} \sim \left|{\mu^2
 \Phi^{N_f-N\over N}\over \Lambda_m^{N_f-3N \over N}}\right|}
For $|\Phi|\gg |\Lambda_m \epsilon^{2N\over N_f-3N}|$ the
correction due to the superpotential \leadpotp\ is more important
than the correction due to the K\"ahler potential \leadpotc.  For
smaller values of $\Phi$ both corrections are negligible.  This
answers our second question.

We conclude that the corrections due to the high energy theory and
other modes at the scale $\Lambda_m$ do not invalidate our
conclusions.  Our perturbative computations in section 2 and the
nonperturbative computations in section 3 are completely under
control and lead to the dominant contributions to the low energy
dynamics.

\newsec{Meta-stable Vacua in SUSY QCD}

In the preceding sections, we have gradually assembled the tools
necessary for analyzing supersymmetry breaking in SQCD. Now let us
put these tools to work. The model of interest is $SU(N_c)$ SQCD
with scale $\Lambda$ coupled to $N_f$ quarks $Q_f$, $\tilde Q_g$,
$f,g=1,\dots,N_f$ (for a review, see e.g.\ \ISrev). We take for
the tree-level superpotential
 \eqn\Wtreesqcd{ W = \Tr\, m M, \qquad\hbox{where}\qquad
M_{fg}=Q_f\cdot \tilde Q_g,} and $m$ is a non-degenerate
$N_f\times N_f$ mass matrix. This theory has $N_c$ supersymmetric
ground states with
 \eqn\sqcdgs{ \langle M\rangle = \left(\Lambda^{3N_c-N_f}\det m
 \right)^{1\over N_c} {1\over m}}
All these supersymmetric ground states preserve baryon number and
correspondingly the expectation values of all the baryonic
operators vanish.

The mass matrix $m$ can be diagonalized by a bi-unitary
transformation. Its diagonal elements can be set to real positive
numbers $m_i$. We will be interested in the case where the $m_i$
are small and of the same order of magnitude. More precisely, we
explore the parameter range
 \eqn\smallmassi{m_i \ll |\Lambda| \qquad ; \qquad {m_i\over m_j}
 \sim 1}
We will consider the cases $N_f>N_c$.  Then, in the limit $m_i\to
0$ with ${m_i\over m_j} \sim 1 $ the expectation values $\langle
M\rangle$ in \sqcdgs\ approach the origin.

The region around the origin can be studied in more detail using
the duality of \NSd\ between our electric $SU(N_c)$ SQCD and a
magnetic  $SU(N_f-N_c)$ gauge theory with scale $\widetilde
\Lambda$, coupled to $N_f^2$ singlets $M_{fg}$ and $N_f$ magnetic
quarks $q_f$ and $\tilde q_f$ in the fundamental and
anti-fundamental representation of $SU(N_f-N_c)$. We will mostly
limit ourselves to the free magnetic range $N_f < {3\over 2}N_c$
where the dual magnetic theory is IR free; higher values of $N_f$
will be briefly discussed at the end of this section. In the free
magnetic range, the metric on the moduli space is smooth around
the origin. Therefore, the K\"ahler potential is regular there and
can be expanded
 \eqn\Kahlersqcd{ K ={1\over \beta} \Tr\,(q^\dagger q+\tilde
 q^\dagger\tilde q)+{1\over \alpha |\Lambda|^2}\Tr\,M^\dagger
 M+\dots,}
where the scale $\Lambda$ appears because the field $M$ is
identified with the microscopic field in \Wtreesqcd, of classical
dimension two.  The dimensionless coefficients $\alpha$ and
$\beta$ are positive real numbers of order one whose precise
numerical values cannot be easily determined because they are not
associated with the holomorphic information in the theory.  Our
quantitative answers will depend on $\alpha$ and $\beta$, but our
qualitative conclusions will not.

The superpotential of the dual $SU(N_f-N_c)$ theory is  \NSd\
 \eqn\Wdualsqcd{ W_{dual} = {1\over \hat \Lambda}\Tr\, M q \tilde
 q +\Tr\,mM. }
The dimensionful coefficient $\hat \Lambda$ is related to the
scales in the problem through \ISrev
 \eqn\lambdaemrel{\Lambda^{3N_c-N_f} \widetilde \Lambda
 ^{3(N_f-N_c)-N_f}=
 (-1)^{N_f-N_c}\hat \Lambda^{N_f} }
The dimensionful parameters of the magnetic theory,
$\widetilde\Lambda$ and $\hat\Lambda$, are not uniquely determined
by the information in the electric theory. This fact is related to
the freedom to rescale the magnetic quarks $q$ and $\tilde
q$.\foot{There is no such freedom to rescale $M$ because it has a
precise normalization in \Wtreesqcd, and correspondingly, we
identify $m$ in the second term in \Wdualsqcd\ with the
microscopic mass matrix $m$.} Rescaling $q$ and $\tilde q$ has a
number of effects. Obviously, it changes the value of $\beta$ in
the K\"ahler potential \Kahlersqcd\ and the value of $\hat\Lambda$
in the superpotential \Wdualsqcd. It also changes the relation
between the electric baryons, $B=Q^{N_c}$ and $\tilde B=\tilde
Q^{N_c}$, and their expressions in terms of the magnetic quarks,
$q^{N_f-N_c}$ and $\tilde q^{N_f-N_c}$.  Finally, $\widetilde \Lambda$
also changes (in
such a way that the relation \lambdaemrel\ is
preserved), because this
rescaling is anomalous under the magnetic gauge group $SU(N_f-N_c)$.

Using the freedom to rescale $q$ and $\tilde q$, we can always set
$\beta=1$, but then we cannot compute both $\hat \Lambda$ and
$\widetilde \Lambda$ in terms of the electric variables.
Alternatively, we can rescale the magnetic quarks to set
$B=Q^{N_c}=q^{N_f-N_c}$ and $\tilde B=\tilde Q^{N_c}=\tilde
q^{N_f-N_c}$. But then we cannot compute $\beta$ (which is
dimensionful). Below, we will find that the two choices are
convenient in different settings.

Let us first consider the case of equal masses, $m_i=m_0$. As
discussed in section 4, the higher order corrections to $K$ in
\Kahlersqcd\ are suppressed by powers of $\Lambda$ and are not
important near $M=q=\tilde q =0$.  Also, $K$ is evaluated at
$m_0=0$; higher order corrections are $\CO({m_0^2\over
\Lambda^2})$ and are negligible. Therefore, the theory based on
the K\"ahler potential \Kahlersqcd\ and the superpotential
\Wdualsqcd\ is the same as the model studied in section 3, with
the parameters and fields related by the dictionary
 \eqn\dictionary{
 \eqalign{&\varphi=q ,\qquad \tilde\varphi=\tilde q,\qquad
 \Phi= {M\over \sqrt{\alpha}\Lambda}, \cr
 & h = {\sqrt{\alpha} \Lambda \over \hat \Lambda},\qquad
 \mu^2 = - {m_0 \hat \Lambda },\qquad
 \Lambda _m= \widetilde \Lambda  , \qquad
 N=N_f-N_c}
 }
Here we have chosen $\beta=1$ and expressed our answers as
functions of $\widetilde\Lambda$ and $\hat \Lambda$.   As a
consistency check, notice that \sqcdgs\ becomes identical to the
supersymmetric vacuum \susymin\ discussed at the end of section 3,
after applying the dictionary \dictionary\ and the identity
\lambdaemrel.

An interesting special case is $N_f=N_c+1$, where the magnetic
gauge group is trivial. Here it is not natural to set $\beta=1$.
Instead, we scale $q$ and $\widetilde q$ such that they are the
same as the baryons $B=Q^{N_c}$ and $\widetilde B=\widetilde
Q^{N_c}$ of the electric theory.   Then, we should replace the
kinetic term for the magnetic quarks in \Kahlersqcd\ with ${1\over
\beta |\Lambda|^{2N_c - 2}}( B^\dagger B + \tilde B^\dagger \tilde
B)$, where again, $\beta$ is a positive dimensionless parameter
which cannot be easily found. The superpotential of the theory is
not that of \Wdualsqcd, but instead, it is \refs{\SeibergBZ,
\ISrev}
 \eqn\WNcpo{W={1 \over \Lambda^{2N_c-1}}(\tilde B^T M B - \det M)
 +\Tr\,mM}
(Note the additional determinant term.) For $N_c>2$ the
determinant interaction is negligible near the origin and this
theory is the same as the $N=1$ version of the theory in section
2.

We can now essentially borrow all our results from
sections 2 and 3. We thus conclude that, for $N_f$ in the range
$N_c+1\le N_f <{3\over 2} N_c$, and for suitable tree-level
quark masses,  SUSY QCD has a meta-stable supersymmetry breaking
ground state near the origin! In fact we have a compact moduli
space of such meta-stable vacua, parameterized by the various
massless Goldstone bosons.

It is surprising that we can establish that a meta-stable state
exists in the strongly coupled region of the theory. Furthermore,
we find the vacuum energy and the entire light spectrum around
that meta-stable state up to two dimensionless numbers $\alpha$
and $\beta$ (or alternatively, $\alpha$ and
$\tilde\Lambda/\hat \Lambda$). Unlike other results in strongly coupled
supersymmetric gauge theory, this result involves also
non-supersymmetric and non-chiral information.

So far, we have derived this result for equal tree-level
quark masses $m_i=m_0\ll |\Lambda|$. But it is straightforward to
generalize to unequal masses $m_i \ll |\Lambda |$.
Consider first the approximation $|m_i -m_0| \ll m_0 \ll |\Lambda|$.
Then, the effect of unequal masses is a small potential of order
$m_i -m_0$ on the moduli space of our meta-stable vacua.  Since
this moduli space is compact, the theory with unequal masses also
has a meta-stable vacuum.

More generally, for arbitrary $m_i \ll |\Lambda |$ we can still use
our low energy effective field theory and conclude that a
meta-stable state exists near the origin. For unequal masses $m_i
\ll |\Lambda |$, the superpotential \Wsunge\ of the macroscopic
theory is replaced with $W_{tree}=h\Tr \varphi \Phi
\tilde\varphi-h\sum _{i=1}^{N_f}\mu _i^2\Phi _{i}^i$, where $\mu
_i^2=-m_i\widehat \Lambda$.  We order the $m_i$ so that $m_1\geq
m_2\dots \geq m_{N_f}>0$. The meta-stable vacuum is then given by
 \eqn\metamineq{\Phi =0, \qquad
 \varphi = \widetilde \varphi ^T= \pmatrix{\varphi _0\cr 0}, \qquad
 \varphi _0={\rm diag}(\mu _1, \mu _2, \dots, \mu_N).}
In this vacuum, the non-vanishing F-terms are $F_{\Phi _{i}^i}$
for $i=N+1, \dots N_f$, and the vacuum energy is $V_0=\sum
_{i=N+1}^{N_f}|h\mu _i^2|$.  For the vacuum \metamineq\ to be
(meta) stable, it is crucial that the $\varphi _0$ expectation
values in \metamineq\ are set by the $N$ largest masses $m_i$.
Replacing one of the $\varphi _0$ entries $\mu _{i\leq N}$ in
\metamineq\ with a $\mu _{i>N}$ does not yield a (meta) stable
vacuum -- the tree level spectrum contains an unstable
mode, sliding down to the vacuum  \metamineq.

What happens for $m_i$ large compared with $|\Lambda|$?  Clearly,
our approximations can no longer be trusted.  In particular,
if all $m_i \gg |\Lambda |$ we have no reason to believe that such a
meta-stable state exists. However, let us try to make one of the
masses, $m_{N_f}$ large while keeping the other masses small.  For
$m_{N_f} \gg | \Lambda |$ we can integrate out the heavy quark and
reduce the problem to that of smaller number of flavors.  As long
as the number of light flavors $\hat N_f$ satisfies $\hat N_f \ge
N_c+1$, our effective Lagrangian argument shows that such a
meta-stable vacuum exists.

Let us try to go one step further and flow down from $N_f=N_c+1
\to N_c$.  We start with $N_c+1$ light flavors with $m_{i=1...N_c}
\ll m_{N_c+1} \ll |\Lambda |$ and find a meta-stable state which
up to symmetry transformations has $B_i=\tilde B_i=0$, for all
$i=1\dots N_c$, and $B_{N_c+1}=\tilde B_{N_c+1}\neq 0$.
 If we can trust this approximation as
$m_{N_c+1} \gg  |\Lambda |$, we find the following picture for the
$N_f=N_c$ problem.  For $m=0$ the low energy theory is
characterized by the modified moduli space of vacua \SeibergBZ
 \eqn\modmod{\det M - B\tilde B = \Lambda^{2N_c}}
and the K\"ahler potential on that space is smooth.  Consider the
theory at the vicinity of the points related to
 \eqn\vicio{M=0 \qquad, \qquad B=\tilde B = i \Lambda^{N_c}}
by the action of the global baryon number symmetry.  The K\"ahler
potential around that point depends on the fields which are
tangent to the constraint \modmod
 \eqn\KahlerNfNc{K={1\over \alpha |\Lambda| ^2} \Tr\, M^\dagger M +
 {|\Lambda| ^2\over \beta } b^\dagger b + \dots}
where $B=i\Lambda^{N_c} e^b$, $\tilde B=i\Lambda^{N_c} e^{-b}$,
and again $\alpha$ and $\beta $ are dimensionless real and
positive numbers which we cannot compute.  Turning on the
superpotential $m_0\Tr\, M$ leaves unlifted, to leading order, the
pseudo-flat directions labelled by $M$ and $b$.  These pseudo-flat
directions are lifted by the higher order terms in \KahlerNfNc\
which we cannot compute. (Note that unlike the case with more
flavors, where the loops of massive but light fields give the
dominant correction to the pseudo-flat directions, here there are
no such light fields which can lead to a reliable conclusion.)
Although we cannot prove it in this case, motivated by the flow
from the problem with one more flavor, we suggest that the
states \vicio\ might also be meta-stable.

So far we have restricted attention to $N_f<{3\over 2}N_c$ where
the magnetic degrees of freedom are IR free.  What happens for
larger values of $N_f$?  Clearly, for $N_f \ge 3 N_c$ the electric
theory is not strongly coupled in the IR and its dynamics is
trivial.  Therefore, our meta-stable states are not present.  For
${3\over 2}N_c <N_f< 3N_c$ the theory flows to a nontrivial fixed
point \NSd.  We can again use the magnetic description which flows
to the same fixed point.  However, the analysis above in the
magnetic theory should be modified in this case.  The duality is
still valid only below $\Lambda$, but unlike the free magnetic
case, here the magnetic theory is interacting in this range. A
closely related fact is that, for nonzero $M$, the dynamically
generated superpotential is \refs{\AffleckMK,\ISrev,\ArgyresCB}:
 \eqn\Wlowa{W_{dyn}  = (N_c-N_f)\left({\det\,M \over \Lambda
 ^{3N_c-N_f}}\right)^{1 \over N_f-N_c}}
(One can check that this is the same as \Wfull\ after using
\dictionary\ and \lambdaemrel.) For $M$ near the origin, this
scales like $ M^{N_f\over N_f-N_c}$ which is larger than $M^3$ and
cannot be neglected in the analysis of the potential.
Equivalently, for these values of $N_f$ and $N_c$ the expectation
value of $M$ \sqcdgs\ is too close to the origin to allow the
existence of our meta-stable state. Finally, the case $N_f={3\over
2} N_c$ is more subtle because the magnetic theory is IR free only
because of its two loop beta function. Here the superpotential
\Wlowa\ scales like $M^3$ and again it cannot be neglected near
the origin.  It is interesting that in this case \Wlowa\ is
independent of $\Lambda$ and in terms of the magnetic variables
the superpotential \Wlowa\ is independent of $\Lambda_m$.

To summarize, we have demonstrated in this section that $SU(N_c)$
SQCD with $N_c+1\le N_f<{3\over2}N_c$ massive flavors exhibits
dynamical meta-stable supersymmetry breaking. In addition, we have
suggested  that the same might be true for $N_f=N_c$. Our calculations
are completely under control when the tree-level masses are in the
regime $m_i \ll |\Lambda |$.  The correction
computed in section 2 due to integrating out light fields is of
order $m_i/|\Lambda |$ and is the leading order correction
to the effective potential.

If we take the masses $m_i$ to all be equal, there is a vector-like $U(N_f)\cong SU(N_f)\times U(1)_B$ global symmetry.  This symmetry is unbroken in the supersymmetric vacua \sqcdgs, which is consistent with their mass gap.  In the meta-stable, dynamical supersymmetry breaking
vacua, the $U(N_f)$ global symmetry is spontaneously broken to $S(U(N_f-N_c)\times U(N_c))$
(plus there is an accidental $U(1)_R$ symmetry).   The meta-stable dynamical supersymmetry breaking vacua is thus a compact moduli space of vacua,
\eqn\mcis{{\cal M}_c\cong {U(N_f)\over S(U(N_f-N_c)\times U(N_c))}.}

Note that there is a bigger configuration space \mcis\ of vacua with broken supersymmetry, versus the isolated supersymmetric vacua.  Perhaps the larger configuration space will favor cosmology initially populating the vacua with broken supersymmetry.

Let us summarize the mass spectrum in the vacua with broken supersymmetry.  There are many heavy states, associated with the microscopic theory, with masses of the order of $\Lambda$.  The fields of the low-energy effective theory are those of the magnetic dual.  Some of these fields get tree-level masses, of the order of $\sqrt{m\Lambda}\ll \Lambda$; this includes the magnetic gauge fields and gauginos, which are Higgsed.  The pseudo-moduli have masses which are smaller, suppressed by a loop factor of the IR free Yukawa coupling of the magnetic dual.  There are massless scalars: the Goldstone bosons of the vacuum manifold \mcis.  There are also massless fermions (including the Goldstino): the $N_c^2$ fermionic partners of the pseudo-moduli $\Phi _0$, i.e. the fermions $\psi _M$ in the null space of both $\ev{q}$ and $\ev{\widetilde q}$.

We also note that the non-trivial topology of the vacuum manifold \mcis\ means that there are
topological solitons, whose lifetime is expected to be roughly the same as
that of the meta-stable vacuum. In 4d, there are $p$-brane topological solitons if
$\pi _{3-p}({\cal M}_c)$ is non-trivial. In particular, the vacuum manifold \mcis\ leads to solitonic strings.

\newsec{$SO(N)$ and $Sp(N)$ Generalizations}

In this section, we give the generalizations of our models to
$SO(N)$ and $Sp(N)$ groups. The $SO(N)$ theory (or more precisely,
$Spin(N)$, so we can introduce sources in the spinor
representation) exhibits a new phenomenon: the meta-stable,
non-supersymmetric vacua are in the confining phase, whereas the
supersymmetric vacua are in a different phase, the oblique
confining phase.  These different phases occur in this case
because the dynamical matter is in an unfaithful representation of
the center of the gauge group, leaving ${\Bbb Z}_2\times {\Bbb
Z}_2$ electric and magnetic order parameters which can not be
screened. The order parameters determine whether Wilson and 't
Hooft loops in the spinor representation of the $SO(N)$ group have
area or perimeter law.  We will argue that, in the meta-stable
vacua with broken supersymmetry, the 't Hooft loop  with
magnetic ${\Bbb Z}_2$ charge has perimeter law, while that with
oblique electric and magnetic ${\Bbb Z}_2$ charges has area law.
In the supersymmetric vacua the situation is reversed: the oblique
charged loop has perimeter law, and the magnetic charged loop has
area law.

\subsec{The $SO(N)$ macroscopic theory}

Consider a model with global symmetry and matter content
\eqn\globalsymson{
 \matrix{ & SO(N) & SU(N_f) & U(1)' & U(1)_R \cr
          & \cr
         \Phi & 1 & \square\square  & -2 &  2\cr
         & \cr
         \varphi & \square & \square  & 1 & 0\cr
         }
 }
The K\"ahler potential is taken to be canonical,
 \eqn\Kahlerson{ K = \Tr\,\varphi^\dagger \varphi+ \Tr\,\Phi^\dagger
  \Phi }
(Because $\Phi$ is a symmetric matrix, the K\"ahler potential has
an extra factor of 2 for the off-diagonal components of $\Phi$.
This will be properly taken into account in the following
analysis.)  The superpotential is taken to be
 \eqn\Wson{ W =
 h\Tr\, \varphi^T \Phi \varphi - h \mu^2\Tr\,\Phi. }
For $\mu \neq 0$, the $SU(N_f)\times U(1)'$ global symmetry is broken to
$SO(N_f)$.

For $N_f>N$ and $\mu\ne 0$, supersymmetry is spontaneously broken
as the rank condition again prevents $F_\Phi$ from all vanishing.
Up to global symmetries, the potential is minimized by
 \eqn\Vminsonii{
 \Phi =\pmatrix{0&0 \cr 0& \Phi_0}, \qquad
 \varphi=\pmatrix{\varphi_0\cr
 0},\qquad {\rm with}\,\,\,\varphi_0^T \varphi_0 = \mu^2\unit_N
 }
where $\Phi_0$ is an arbitrary  $(N_f-N)\times(N_f-N)$ symmetric
matrix, and $\varphi_0$ is an $N\times N$ matrix subject to the
condition in \Vminsonii. All vacua on this space of classical
pseudo-flat directions have degenerate vacuum energy density
 \eqn\Vminson{
 V_{min} = (N_f-N)|h^2\mu^4|.
 }
We can use the $SU(N)$ result of section 2 to show that \Vminson\
is indeed the absolute minimum of the potential. The classical
potential of this $SO(N)$ theory satisfies $V_{SO(N)}\geq
|h\varphi\varphi^T-h\mu^2|^2 \geq (N_f-N)|h^2\mu ^4|$, where for
the first inequality we simply set $\Phi =0$ and in the second we
used the $SU(N)$ result, restricted to the smaller space where
$\tilde\varphi=\varphi ^T$.

We now show that perturbative quantum effects lift the above classical
vacuum degeneracy, and that a local minimum of the one-loop
effective potential is (up to symmetries)
 \eqn\maximalson{\Phi_0=0,
\qquad \varphi_0=\mu\unit_{N}.}
Of the classical vacua \Vminsonii, this has maximal unbroken
global symmetry, with $SO(N)\times SO(N_f)\times U(1)_R\rightarrow
SO(N)_D\times SO(N_f-N)\times U(1)_R$. We will focus on the
leading perturbative corrections to the effective potential,
expanded around the vacuum \maximalson.

Expanding around \maximalson, we write the fields as
\eqn\mqfluctsonge{
\Phi =\pmatrix{\delta Y& \delta Z^T\cr \delta Z& \delta\hat\Phi},
\qquad \varphi=\pmatrix{\mu+\delta\chi_A+\delta\chi_S\cr
\delta\rho}.
 }
where $\delta\chi_A$ and $\delta\chi_S$ denote the antisymmetric
and symmetric part, respectively, of $\delta\chi_A+\delta\chi_S$.
The Goldstone bosons of the broken global symmetry are ${\rm
Re}\left({\mu ^*\over |\mu|}\delta\chi_A\right)$ and ${\rm
Re}\left({\mu ^*\over |\mu|}\delta\rho\right)$. The former are in
the adjoint of $SO(N)\times SO(N)_F/SO(N)_D\cong SO(N)$ (with
$SO(N)_F\subset SO(N_f)$), and hence they are antisymmetric; the
latter are in $SO(N_f)/SO(N)_F\times SO(N_f-N)$.

There are also the classically massless pseudo-moduli fields,
\eqn\pmfson{
\delta\hat\Phi\,\,\, {\rm and}\,\,\, \delta\hat\chi\equiv {\rm
Im}\left({\mu ^*\over |\mu|}\delta\chi_A\right).
 }
These are lifted at one-loop, with an effective potential that is
constrained by the symmetries and dimensional analysis to have the
form
 \eqn\Vonegenson{ V_{eff}^{(1)} =\left| h^4\mu^2\right|\left({1\over2} a\,\Tr\,\delta\hat\chi^T\delta\hat\chi +
 b\, \Tr\,\delta\hat\Phi^\dagger \delta\hat\Phi\right) + \dots
  }
for some numerical coefficients $a$ and $b$.   These coefficients
are computed in appendix B; the calculation is very similar to the
$SU(N)$ case. The result is
 \eqn\Vonegensonii{ V_{eff}^{(1)} =
 {|h^4 \mu^2|(\log 4-1)\over 2\pi^2}\left( (N_f-N)
 \Tr\, \delta\hat\chi^T\delta\hat\chi +
 N\,\Tr\,\delta\hat\Phi^\dagger \delta\hat\Phi\right) + \dots
 }
The mass-squares of the pseudo-moduli are positive. The one-loop
potential \Vonegensonii\ stabilizes all (non-Goldstone-boson)
pseudo-flat directions at the origin, with mass $\sim |h^2 \mu|$.

Now consider the effect of gauging $SO(N)$, taking it to be
infrared free, $N_f>3(N-2)$.  Then this theory becomes a
macroscopic, low-energy effective theory, valid for energies below
some cutoff scale $\Lambda _m$.  The vacuum is still \maximalson\
since the D-terms vanish there. The gauge group is completely
broken in the vacuum, so the $SO(N)$ vector bosons, together with
the pseudo-moduli and Goldstone bosons derived from
$\delta\chi_A$, acquire masses from the super-Higgs mechanism.
Exactly as in the $SU(N)$ case, at leading order the tree-level
$SO(N)$ vector supermultiplet masses are not split by the
supersymmetry breaking. Thus, the one-loop potential for the
remaining pseudo-modulus $\delta\hat\Phi$ is the same as in
\Vonegenson.

Evidently, gauging $SO(N)$ does not significantly affect the
non-supersymmetric vacuum \maximalson. But just as for $SU(N)$, it
does introduce a supersymmetric vacuum elsewhere in field space.
Giving $\Phi$ general, non-zero expectation values in \Wson\ gives
the fields $\varphi$ masses, and integrating them out leads to the
low-energy effective superpotential
 \eqn\wduallso{W_{low}=(N-2)\left( h^{N_f} \Lambda _m ^{3(N-2)-N_f}\det \Phi
\right)^{1/(N-2)}-h\mu ^2\Tr\, \Phi,} where the first term arises
from gaugino condensation in the low-energy $SO(N)$ Yang-Mills
theory, with scale related to $\Lambda _m$ by matching at the
scale $\langle h\Phi\rangle$ where $\varphi$ are integrated out.
The first term in \wduallso\ leads to dynamical supersymmetry
restoration, with $N_f-N+2$ supersymmetric vacua at
 \eqn\wdualsosv{\ev{\Phi}={\Lambda _m\over h}\epsilon
 ^{2(N-2)/(N_f-N+2)}
\unit_{N_f}, \qquad\hbox{where}\qquad \epsilon \equiv {\mu \over
\Lambda _m}.} Again, we take $|\epsilon |\ll 1$ parametrically small
to be able to reliably compute within the macroscopic effective
theory.  We will see that this also ensures that the meta-stable,
non-supersymmetric vacuum \maximalson\ is long lived.

\subsec{The ultraviolet theory: $SO(N_c)$ with $N_f< {3\over
2}(N_c-2)$ massive flavors}

The macroscopic theory of the previous subsection is the infrared
free dual \IntriligatorID\ of $SO(N_c)$ with $N_f<{3\over
2}(N_c-2)$ massive flavors, which is asymptotically free. The
dictionary relating the microscopic $SO(N_c)$ theory to the
macroscopic $SO(N)$ theory \Wson\ is much as in \dictionary,
except that here $N=N_f-N_c+4$, and there are no $\tilde\varphi$
or $\tilde q$ fields. The supersymmetric vacua \wdualsosv\ are
those discussed in \IntriligatorID, and expected from the Witten
index of $SO(N_c)$ with massive matter.

There are some special cases in the duality of \IntriligatorID.
For $N_f=N_c-2$, the magnetic dual is $N=2$, i.e. $SO(2)$; the
infrared theory is then in the Coulomb phase.   The theory \Wson\
describes the $N_f$ magnetic monopoles, $\phi ^i$,  near
$M_{ij}=Q_i\cdot Q_j=0$. Actually, as mentioned in
\IntriligatorID, the superpotential \Wson\ in this case should be
multiplied by a holomorphic function $f(t)$, with $t=\det
M/\Lambda^{2(N_c-2)}$ and $f(0)=1$.  The leading order mass
spectrum of the meta-stable, supersymmetry breaking vacuum
involves only $f(0)$, and so it is completely independent of this
function.

In the vacuum \Vminsonii, the magnetic $SO(2)$ is Higgsed, and the
unbroken electric $SO(2)$ is confined.   For $N_f>2$, these vacua
break supersymmetry, and are meta-stable.  The supersymmetric
vacua of the electric theory with massive flavors comes from the
massless dyon point of \IntriligatorID, at $\det M_{ij} = 16
\Lambda^{2N_c-4}$; upon adding masses for the electric flavors,
these dyons condense, and lead to the supersymmetric vacua
\wdualsosv. Condensing of the dyons leads to oblique confinement.
We thus find that our meta-stable non-supersymmetric vacuum, and
the supersymmetric vacua, are in different phases: confining, and
oblique confining, respectively.  Wilson and 't Hooft loops in the
spinor representation can not be screened by the dynamical matter,
so we have ${\Bbb Z}_2\times {\Bbb Z} _2$ order parameters which
can distinguish between the confining and oblique confining
phases.  The loop with area law in the
non-supersymmetric vacuum will have perimeter law in the
supersymmetric vacua, and vice versa.  We expect that this is also true
for  $N_f>N_c-2$, because we do not expect a phase transition if we
give some flavors large masses,
and flow down to $N_f=N_c-2$.

For $N_f=N_c-3$, there are two physically inequivalent phase
branches \IntriligatorID.  The supersymmetric vacua of the theory
with mass terms come from the branch with a dynamical
superpotential $W_{dyn}\sim 1/\det M$.  The other branch has the
fields of \Wson\ with $N=1$, where $SO(1)$ means that there is one
magnetic color index, but no corresponding gauge group.  The
superpotential \Wson\ can in general be modified by multiplying it
by a holomorphic function $f(t)$, with $t=\det M
(M^{ij}q_iq_i)/\Lambda ^{2N_c-3}$, with $f(0)=1$ \IntriligatorID.
This branch leads to our meta-stable non-supersymmetric vacua,
with a spectrum that is independent of the function $f(t)$.

For $N_f=N_c-4$, there are again two physically inequivalent
branches, one with dynamical superpotential and one with
$W_{dyn}=0$ \IntriligatorID.  The branch with dynamical
superpotential leads to the expected supersymmetric vacua upon
adding mass terms for the flavors of the microscopic theory.  On
the other hand, the vacua with $W_{dyn}=0$ break supersymmetry
upon adding $W_{tree}=m\Tr M\equiv -h\mu ^2 \Tr \Phi$
\IntriligatorID.  The leading K\"ahler potential near the origin
is
 \eqn\korigex{K=\Tr \Phi ^\dagger \Phi +{c\over |\Lambda
 _m|^2}\Tr (\Phi ^\dagger \Phi )^2+\dots,}
with $c$ a number of order one. If $c$ is negative (positive), the
potential at the origin curves up (down). On the other hand, for
large $\Phi$ the scalar potential must curve up, because there the
K\"ahler potential must agree with the classical K\"ahler
potential of the electric description, $K\sim \sqrt{M^\dagger M}$,
with $M\sim \Phi$. So assuming that the K\"ahler potential is
nondegenerate for all $\Phi$, we conclude that there must exist a
non-supersymmetric vacuum, {\it somewhere} on the $W_{dyn}=0$
branch, regardless of the sign of $c$. This non-supersymmetric
vacuum is stable on this branch of the theory; it can only decay
via tunnelling to the $W_{dyn}\ne 0$ branch. The situation here
should be compared with the analogous situation in $SU(N_c)$ SQCD
with $N_f=N_c$. There we have only one branch, and therefore we
cannot conclude definitively that there is a meta-stable vacuum.
But the analogy with this $SO(N_c)$ example further motivates our
suggestion above that such a meta-stable state exists.

\subsec{$Sp(N)$ Theories}

The $Sp(N)$ theory is especially simple.  It does not have the
richness of the different phases of the $SO(N)$ theory and it does
not have the baryons of the $SU(N)$ theory.   Our conventions are
such that $Sp(N)$ consists of all $A\in SU(2N)$ satisfying $A^T
J_{2N} A=J_{2N}$, with $J_{2N}=\unit_{N}\otimes (i\sigma _2)$. In
particular,  $Sp(1)\cong SU(2)$.

The macroscopic, low-energy theory has symmetries and matter
content \eqn\globalsymspn{
 \matrix{ & Sp(N) & SU(2N_f) & U(1)'&U(1)_R \cr
          & \cr
         \Phi & 1 & \doub  & -2&2\cr
         & \cr
         \varphi & \square & \overline{\square}  & 1&0\cr
         }
}
canonical K\"ahler potential, and superpotential
 \eqn\Wspn{
 W = h\Tr\, \varphi^T\Phi \varphi J_{2N}   - h \mu^2\Tr\,\Phi
 J_{2N_f}.
 }
For $\mu \neq 0$, $SU(2N_f)\times U(1)'$ is broken to $Sp(N_f)$.  We take
$N_f>3(N+1)$, so the $Sp(N)$
gauge coupling is infrared free.  Again, we first consider the
theory for zero $Sp(N)$ gauge coupling.

The scalar potential has an absolute minimum, with energy density
 \eqn\Vminspn{V_{min} = 2(N_f-N)|h^2\mu^4|.}
Indeed, we have $V_{Sp(N)}\geq \left| h(\varphi J_{2N}\varphi ^T
 -\mu ^2J_{2N_f})\right|^2 \geq 2(N_f-N)\left| h^2\mu ^4\right|$,
 where for the first inequality we sent $\Phi =0$, and
in the second we used the result for $SU(2N)$, with $2N_f$
flavors, restricted to a smaller subspace where $\widetilde
\varphi =J_{2N}\varphi ^T$. Up to unbroken global symmetries, the
classical vacua with this minimum energy are
\eqn\Vminspnii{
 \Phi =\pmatrix{0 & 0 \cr 0  & \Phi_0},
 \qquad \varphi=\pmatrix{\varphi_0
 \cr  0},\qquad {\rm with}\,\,\,\,\,  \varphi_0 J_{2N} \varphi_0^T = \mu^2 J_{2N}
}
where $\Phi_0$ is an arbitrary $2(N_f-N)\times 2(N_f-N)$
antisymmetric matrix, and $\varphi_0$ is a $2N\times 2N$ matrix.

The one-loop effective potential lifts this classical vacuum
degeneracy, and the local minimum is at the point of maximal
unbroken global symmetry:
 \eqn\maximalspn{
 \Phi_0=0,\qquad \varphi_0=\mu\unit_{2N}
 }
which leaves unbroken a $Sp(N)_D\times Sp(N_f-N)\times U(1)_R$
global symmetry. Let us decompose the fluctuations around this
point as
 \eqn\fluctspn{
\Phi=\pmatrix{ \delta Y & \delta Z^T\cr -\delta Z &
\delta\hat\Phi},\qquad \varphi=\pmatrix{ \mu + J_{2N}(
\delta\chi_{A}+\delta\chi_{S})\cr \delta\rho}
 }
where again by $\delta\chi_{A}$ and $\delta\chi_{S}$ we mean the
antisymmetric and symmetric part, respectively, of
$\delta\chi_A+\delta\chi_S$. The Goldstone bosons of the broken
global symmetry are
\eqn\spgb{
{\mu ^*\over |\mu|}\delta\chi_{S}-{\mu \over
|\mu|}J_{2N}\delta\chi_{S}^* J_{2N}\,\,\,\, {\rm and}\,\,\,\, {\mu
^*\over |\mu|}\delta\rho+{\mu \over |\mu|}J_{2(N_f-N)}\delta\rho^*
J_{2N}
}
The former are in the adjoint of $Sp(N)\times Sp(N)_F/Sp(N)_D\cong
Sp(N)$, and hence symmetric; the latter are in
$Sp(N_f)/Sp(N)_F\times Sp(N_f-N)$.  There are also classically
massless pseudo-moduli fields
\eqn\sppf{
\delta\hat\Phi\,\,\,\, {\rm and}\,\,\,\, \delta\hat\chi\equiv
{\mu^*\over |\mu|}\delta\chi_{S} +
{\mu\over|\mu|}J_{2N}\delta\chi_{S}^* J_{2N}
}

Once again, the global symmetries and dimensional analysis
constrain the one-loop effective potential to have the form
 \eqn\Vonegenspn{ V_{eff}^{(1)} =
 \left| h^4\mu^2\right|\left({1\over2} a\,\Tr\,(J_{2N}\delta\hat\chi)^2+
 b\, \Tr\,\delta\hat\Phi^\dagger \delta\hat\Phi\right) + \dots
  }
The coefficients $a$ and $b$ are computed in appendix B, and the
result is
\eqn\Vonespnii{ V_{eff}^{(1)} = {|h^4\mu^2|(\log 4-1)\over \pi^2}\left(
 {1\over4}(N_f-N)\Tr (J_{2N}\delta\hat\chi)^2
 +N\Tr\,\delta\hat\Phi^\dagger \delta\hat\Phi\right). }
The pseudo-flat directions are thus stabilized, with non-tachyonic masses $\sim |h^2
\mu|$.

As in the $SU(N)$ and the $SO(N)$ examples, gauging $Sp(N)$ does
not affect the one-loop potential \Vonespnii, because the
classical masses of the (completely Higgsed) $Sp(N)$ vector
multiplet are supersymmetric. And, as above, gauging $Sp(N)$ leads
to superymmetric vacua,  by dynamical supersymmetry restoration,
elsewhere in field space.  The supersymmetry-breaking vacua
\maximalspn\ are thus meta-stable.

The theory \Wspn\ is the dual of a microscopic theory given by
$Sp(N_c)$ gauge theory, with $2N_f$ fundamental flavors $Q$
\IntriligatorNE.   The dictionary is much as in \dictionary,
except that $N=N_f-N_c-2$.  For $N_f<{3\over 2}(N_c+1)$, the
macroscopic $Sp(N)$ theory \Wspn\ is infrared free.  Our analysis
of the macroscopic theory shows that the microscopic $Sp(N_c)$
theory, with small masses for the fundamental flavors, has the
meta-stable, supersymmetry breaking  vacua, given by \Vminspnii\
and \maximalspn.

For $N_f=N_c+2$, we have $N=0$, so the dual theory does not
include gauge fields. The fields of the low-energy theory are just
$M$, with a superpotential \IntriligatorNE,
 \eqn\WNspo{W=-{{\rm Pf}M\over \Lambda ^{2N_c+1}}+\Tr mM.}
In many respects this case is similar  to the $SU(N_c=N_f-1)$
theories.   However, unlike these theories the superpotential
\WNspo\ does not include cubic terms (for $Sp(N_c>1)$). and
therefore only $\Tr mM$ is important near the origin.  Then,
depending on the K\"ahler potential, this term could lead to a
supersymmetry breaking meta-stable state near the origin.  In this
respect this situation is similar to the $SU(N_c=N_f)$ theories.

Finally, we can analyze the case where one mass eigenvalue is much
larger than the others, flowing to $N_f=N_c+1$, where there is a
quantum modified moduli space constraint \IntriligatorNE,
analogous to that of $SU(N_c=N_f)$ SQCD.  The analysis of the
theory with mass terms in analogous to the discussion following
\modmod, with some components of $M$ here playing the role of the
baryon expectation values in \vicio\ ($Sp(N)$ does not have
baryons). Again, we suggest here that for $N_f=N_c+1$ and small
tree-level masses, $Sp(N_c>1)$ SQCD has meta-stable
supersymmetry-breaking vacua near the origin of field space.

\newsec{Estimating the Lifetime of the Meta-stable Vacua}

In this section, we will show that our meta-stable,
non-supersymmetric vacua can be made parametrically long lived, by
taking the parameter $\epsilon \equiv \mu /\Lambda _m\sim
\sqrt{m/\Lambda}$ to be sufficiently small.  We ignore quantum
gravity effects,\foot{If we add a constant superpotential, so that the meta-stable vacuum has our observed vacuum energy, then the supersymmetric vacua are anti-deSitter.  This can lead to a
suppressed quantum gravity tunneling rate.}
and consider only the
semi-classical field theory decay as in \ColemanPY. The
semi-classical decay probability, which sets the decay rate, is
given by $\exp(-S)$, where $S$ is the ``bounce" action (the
difference between the Euclidean action of the tunneling
configuration and that of remaining in the meta-stable vacuum),
times an irrelevant one-loop prefactor. We will argue that $S$ is
parametrically large as $\epsilon \rightarrow 0$, making the
lifetime arbitrarily long.

In order to give a qualitative estimate of the bounce action $S$,
we need to give a qualitative picture of the potential for the
scalar fields, $\Phi$ and $\varphi$ and $\widetilde \varphi$.
Recall that our meta-stable non-supersymmetric vacuum is (we
discuss the $SU(N)$ case; the discussion for $SO(N)$ and
$Sp(N)$ is completely analogous):
 \eqn\metarev{\Phi =0, \qquad \varphi = \widetilde \varphi ^T
 =\pmatrix{\mu \unit_N\cr 0}, \qquad V_+=(N_f-N)\left| h^2\mu ^4\right| .}
The supersymmetric vacuum \susymin\ on the other hand has
 \eqn\susyminr{\Phi={\mu \over h}{1\over \epsilon
 ^{(N_f-3N)/(N_f-N)}}\unit_{N_f}, \qquad \varphi =\widetilde
 \varphi =0, \qquad V_0=0.}
Because we take $N_f> 3N$, which is the condition for the
macroscopic theory to be infrared free, the supersymmetric minimum
\susyminr\ is parametrically far away from the meta-stable
non-supersymmetric vacuum \metarev\ as $\epsilon \rightarrow 0$.
As we shall see, this large distance $\Delta \Phi$ in field space
guarantees a parametrically large bounce action $S$.

The bounce action is expected to come from the path in field space
with the least potential barrier between the vacua \metarev\ and
\susyminr.   Computing the classical potential from \Wsunge, we
find terms $V_{cl}\supset \left| h\varphi\Phi \right| ^2+\left|
h\Phi \tilde \varphi \right| ^2$, which provide a large potential
energy cost to having both $\Phi$ and $\varphi$ or $\tilde\varphi$
being non-zero.  The most efficient path is thus to climb quickly
from \metarev\ up to a point near the local peak \eqn\peak{\Phi
=0, \qquad \varphi = \widetilde \varphi =0, \qquad V_{peak}=N_f
\left| h^2\mu ^4\right| .} {}From there, we can take the path of
increasing $\Phi$, toward the minimum \susyminr, keeping $\varphi
=\tilde \varphi =0$; the potential along this path is extremely
flat, as $\epsilon \rightarrow 0$, sloping only very gently\foot{This
gentle slope could be also useful for inflation or quintessence.} toward
the minimum \susyminr. A schematic picture of the potential is
shown in fig.\ 1.

\bigskip
\centerline{\epsfxsize=0.70\hsize\epsfbox{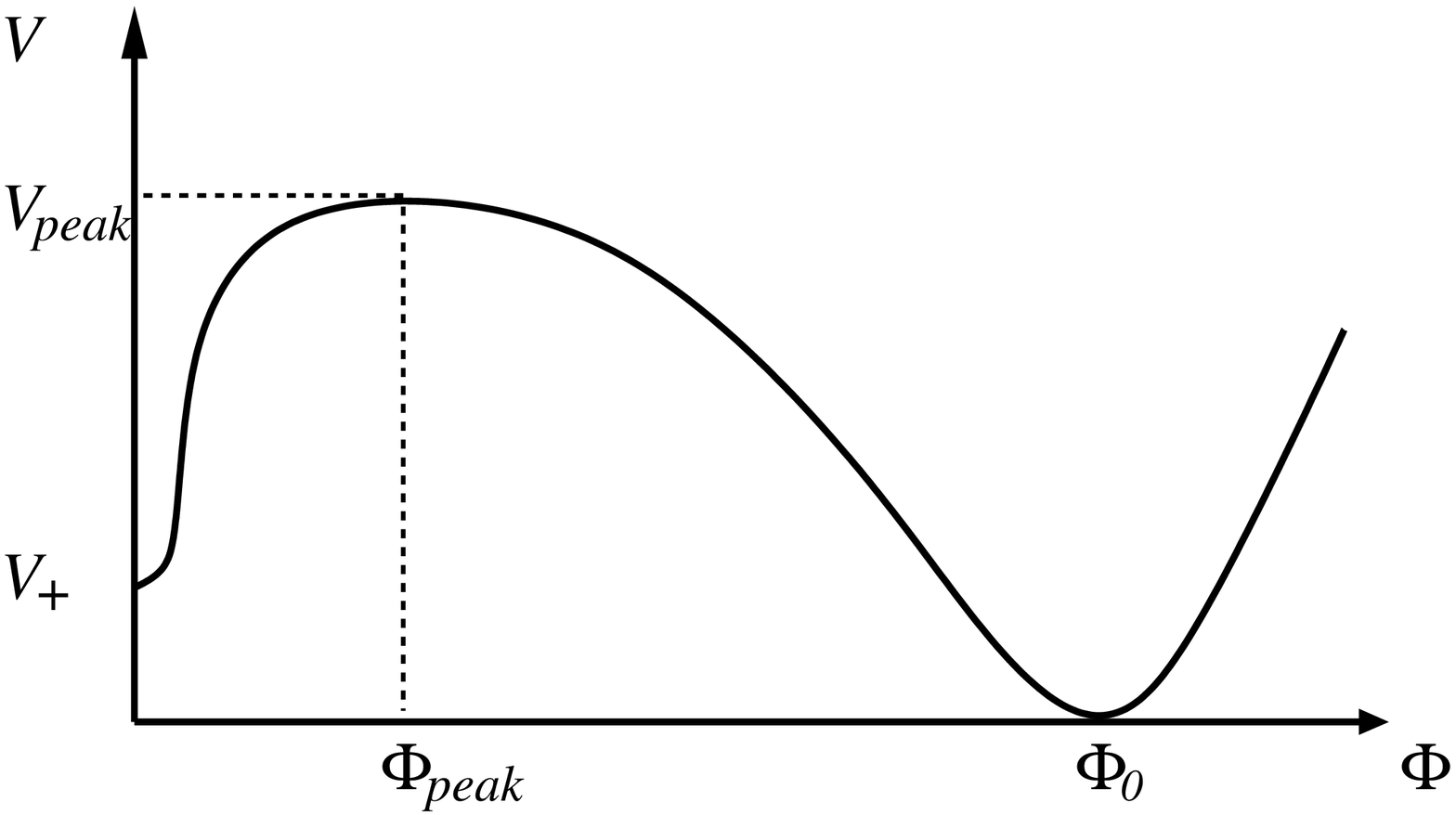}}
\noindent{\ninepoint\sl \baselineskip=8pt {\bf Figure 1}:{\sl $\;$
The potential along the bounce trajectory.  The peak is at
$\Phi_{peak} \sim \mu$ and the supersymmetric minimum with
vanishing potential is at large field $\Phi_0 \sim
\mu/\epsilon^{(N_f-3N)/(N_f-N)} \gg \mu$.  The values of the
potential at the local minimum $V_{+} $ and at the peak $V_{peak}$
are of order $\mu^4$.}}
\bigskip

The thin wall approximation \ColemanPY\ is not appropriate for
computing the bounce action of such a potential.  The needed
calculation of the bounce action can be modelled by a triangle
potential barrier.  Then, using the results of \DuncanAI\ we find
 \eqn\bounceS{S\sim {({\Delta \Phi})^4\over V_+}\sim {1\over
 |\epsilon |^{4(N_f-3N)/(N_f-N)}} \gg 1.}
Taking $\epsilon \rightarrow 0$, we can make the minimal bounce
action arbitrarily large, and thus make the meta-stable vacuum
arbitrarily long lived.

It is amusing to consider the very different magnification scale
of the potential in the microscopic description of the theory and
in the macroscopic description.  The relation \bounceS\ applies in
both descriptions.  In the macroscopic description, we have
$\epsilon = \mu /\Lambda _m$, with $\mu$ held fixed and  the
cutoff scale $\Lambda _m \rightarrow \infty$.  Here the large
action \bounceS\ is intuitive: the vacua \metarev\ and \susyminr\
appear widely separated in field space. On the other hand, in the
microscopic description, we have $\epsilon \sim \sqrt{m/\Lambda}$,
and we hold $\Lambda$ fixed and take $m$ to zero.  Here we are
looking at the potential with a very different magnification
scale, and the parametrically large action is less intuitive: the
vacua \metarev\ and \susyminr\ appear as tiny features, two close
vacua separated by a tiny barrier. Nevertheless, the bounce action
only depends on the ratio $\epsilon$, not the overall scale $\mu$,
so the expression \bounceS\ remains valid.  The decay rate of the
meta-stable vacuum can be made exponentially parametrically small,
by taking $\epsilon$ sufficiently small, whether we are in the
macroscopic scaling where the features of the potential appear
large, or in the microscopic scaling where they appear small.

\newsec{Preliminary Thoughts about Model Building}

This work was motivated by attempts to find new models of
supersymmetry breaking and new mechanisms to communicate
supersymmetry breaking to the Standard Model. We hope that the
theories studied in this paper are a modest step towards building
a simpler and more elegant model of dynamical supersymmetry
breaking. Of course, many challenges lie ahead, and we have not
succeeded in overcoming these challenges. But we would like to
share some of our preliminary ideas about them.

{\it (1) Naturalness.}  The small parameter which controls our
approximations is $\epsilon \sim \sqrt{m/\Lambda}$ and the vacuum
energy\foot{Of course, the actual vacuum energy density includes a
negative supergravity contribution from the value of the
superpotential in the minimum.} is proportional to
$|m^2\Lambda^2|$. Since it is proportional to a power of
$\Lambda$, it is nonperturbative.  However, since it is also
proportional to the tree level parameter $m$, our model does not
satisfy the purist's requirement that all low energy scales are
dynamically generated.   Therefore, we would like to find other
theories, using the same ideas as in our models, where the role of
the parameter $m$ is played by some marginal or irrelevant
coupling constants. For example, we can imagine that the
microscopic theory has such an operator suppressed by a power of
the Planck scale (or some other high energy scale), ${\lambda\over
M_p^\Delta }\CO$ with $\lambda\sim 1$. If this operator acquires a
dynamical F-term $F_{\CO}\sim\Lambda^{2+\Delta}$, then the vacuum
energy is of order $\lambda^2 \Lambda^{4+2\Delta} \over
M_p^{2\Delta}$. This way supersymmetry is broken at a naturally
small scale.\foot{Other models which are worth exploring are based
on similar dualities, e.g.\ those of
\refs{\DKi\DKAS\DKNSAS-\BrodieVX}. These theories contain many
operators ${\cal O}$ with large dimension $\Delta _0$ at weak
gauge coupling, but dimension $\Delta =1$ in the infrared, where
they are free.  Adding them to the superpotential could lead to
meta-stable, non-supersymmetric vacua, by an argument completely
analogous to our rank condition in the dual theory.   A more
detailed analysis is needed to determine whether the interacting
sector of the infrared theory changes our conclusions about the
meta-stable non-supersymmetric vacuum.}

{\it (2) Direct mediation.} A longstanding goal of SUSY
phenomenology, first discussed in \AffleckXZ\ and later analyzed
by various authors (see e.g.\ \LutyVR\ and references therein), is
to find a simple model of {\it direct mediation} of supersymmetry
breaking in which the standard model gauge group couples directly
to the supersymmetry breaking sector. The basic idea of direct
mediation is that the supersymmetry breaking sector has a large
global symmetry $G$ and a subgroup of it $H\subset G$ is gauged
and is identified with (part of) the standard model gauge group.
One of the hallmarks of our theories is that they have large
global symmetries $G$ which could be used this way.

Consider, for example, gauging the $SU(N_f)$ symmetry of our SUSY
QCD example in section 3.  Then, the gauge group below the scale
$\Lambda$ is $SU(N_f-N_c) \times SU(N_f)$ where the $SU(N_f-N_c)$
gauge theory is dual to the microscopic $SU(N_c)$ theory.  In our
meta-stable vacuum this symmetry is broken $ SU(N_f-N_c) \times
SU(N_f) \to SU(N_f-N_c) \times SU(N_c)$ where the first factor is
embedded diagonally in $SU(N_f-N_c) \times SU(N_f) $, and the
second factor is a subgroup of $SU(N_f)$.  It is interesting that
some of the low energy gauge fields are partially electric and
partially magnetic.  In the context of direct mediation of
supersymmetry breaking we can think of this low energy gauge group
(or a subgroup of it) as included in the standard model. Clearly,
depending on the details of such a construction, we might need to
abandon simple unification.

An obstacle for direct mediation is that, if we identify a
subgroup of the standard model, e.g.\ the color $SU(3)_c$ symmetry
with a subgroup $H$ of the flavor symmetry $G$ of the
supersymmetry breaking sector, the colors of that sector lead to
additional $SU(3)$ flavors. If there are too many such flavors,
$SU(3)_c$ can have a Landau pole at a dangerously low scale.   We
do not have a solution to this problem.  But we would like to
suggest that the theory viewed at low energies as $SU(3)$ could be
related in a complicated way to a more microscopic gauge symmetry.
(In the particular example of the previous paragraph, however,
this does not actually help.)

{\it (3) R-symmetry problem.} Models of dynamical supersymmetry
breaking with no supersymmetric vacua must either have a
non-generic superpotential, or must have global $U(1)_R$ symmetry
\refs{\AffleckXZ,\NelsonNF}. However, in order to have nonzero
Majorana gluino masses this R-symmetry should be broken, and to
avoid a massless Goldstone boson this R-symmetry should be
explicitly broken. This explicit breaking could restore
supersymmetry. The authors of \BaggerHH\ pointed out that this
problem can be solved using gravitational interactions. In our
theories there is no exact R-symmetry and hence there exist
supersymmetric vacua. But the existence of an accidental
R-symmetry near the origin leads to a supersymmetry breaking
meta-stable state. The small effect of the explicit $U(1)_R$
breaking in this meta-stable state might be strong enough to avoid
the R-symmetry problem.\foot{M.~Dine has pointed out to us that as
it stands our theory has a discrete R-symmetry which prevents
gluino masses.  However, such a symmetry can be explicitly broken,
e.g.\ by adding nonrenormalizable baryon operators in the
microscopic superpotential.}

It is interesting to compare our models with the discussion of
``R-color" in \AffleckXZ, which is a non-Abelian gauge theory that
was introduced in order to explicitly break the $U(1)_R$ symmetry.
Our models in section 3 fit that pattern.  The theory of section 2
has  an R-symmetry and it breaks supersymmetry with a stable
minimum.  The $SU(N)$ gauge interactions added in section 3
explicitly break the R-symmetry, and they also introduce a
supersymmetric state far in field space.  Such supersymmetry
restoration is a common phenomenon with R-color and was often
considered a problem.  However, the microscopic theory of section
5 gives another perspective on the issue.  Here the R-symmetry is
broken in the $SU(N_c)$ microscopic theory.  In the meta-stable
state, the $SU(N_c)$ gauge interactions dynamically break
supersymmetry, and they also break the R-symmetry.  The role of
R-color is played by their magnetic dual, the $SU(N)$ gauge
fields.

\bigskip

\noindent {\bf Acknowledgments:}

We would like to thank T.~Banks, M.~Dine, J.~Maldacena and
S.~Thomas for useful discussions. The research of NS is supported
in part by DOE grant DE-FG02-90ER40542. The research of DS is
supported in part by a Porter Ogden Jacobus Fellowship and by NSF
grant PHY-0243680. The research of KI is supported in part by UCSD
grant DOE-FG03-97ER40546 and by the IAS Einstein Fund; KI would
like to thank the IAS for their hospitality and support on his
sabbatical visit.   We would like to dedicate this paper to the
memory of John Brodie.  Any opinions, findings, and conclusions or
recommendations expressed in this material are those of the
author(s) and do not necessarily reflect the views of the National
Science Foundation.

\appendix{A}{F-term Supersymmetry Breaking}

\subsec{Generalities}

Spontaneous supersymmetry breaking requires an exactly massless
Goldstino fermion $\psi _X $.  In simple models it originates from
a chiral superfield $X$. The scalar component $X$ can get a mass
{}from either non-canonical K\"ahler potential terms, or more
generally from corrections to the $X$ propagator from loops of
massive fields. Consider,  a theory of a single chiral superfield
$X$, with linear superpotential with coefficient $f$ (with units
of mass${}^2$), \eqn\wlin{W =f X,} and effective K\"ahler
potential $K(X, X^\dagger)$.  Supersymmetry is spontaneously
broken by the expectation value of the F-component of $X$.  The
potential, $V=K_{XX^\dagger}^{-1}|f |^2$, is non-vanishing as long
as the K\"ahler metric is non-singular. The fermion $\psi _X$ is
the exactly massless Goldstino. If $K=K_{can}=XX^\dagger$, then
the scalar component of $X$ is also massless; the potential is
$V=|f|^2$, independent of $\ev{X}$, so there are classical vacua
for any $\ev{X}$.  This vacuum degeneracy is lifted by any
non-trivial K\"ahler potential. For example, if near the origin
$K=XX^\dagger -{c\over |\Lambda |^2}(XX^\dagger)^2+\dots$, then
there is a stable supersymmetric vacuum at the origin if $c>0$. In
this vacuum, the scalar component of $X$ gets mass $m_X^2\approx
4c|f |^2/|\Lambda |^2$. If $c<0$, the origin is not the minimum of
the potential.

The macroscopic, low-energy effective field theory must be under
control to determine whether or not supersymmetry is broken. For
example, $SU(2)$ with an $I=3/2$ matter field has an effective low
energy superpotential \wlin.  If the low energy theory is a free
theory of a composite field $X$, as is suggested by non-trivial 't
Hooft anomaly matching, supersymmetry is spontaneously broken.  If
instead the low energy theory is an interacting conformal theory,
supersymmetry is unbroken \ISS.

In the example \wlin, a singularity in the K\"ahler metric signals
the need to include additional light degrees of freedom. Suppose
that an additional field $q$ becomes massless at a particular
value of $X$, which we can take to be $X=0$, so
 \eqn\wlinq{W=hXqq+fX.}
For $f =0$, there is a moduli space of supersymmetric vacua,
labelled by $\ev{X}$, and $q$ can be integrated out away from the
origin. Turning on $f $ lifts this moduli space, but the theory no
longer breaks supersymmetry, as there is a supersymmetric vacuum
at  $X=0$, $q=\sqrt{-f/h}$. To determine whether or not
supersymmetry is broken requires that the macroscopic low-energy
theory be correctly identified.

In this paper, we will be interested in the one-loop effective
potential for pseudo-moduli (such as $X$), which comes from
computing the one-loop correction \CWgen\ to the vacuum energy. In
\CWgen, ${\cal M}^2$ stands for the classical mass-squareds of the
various fields of the low-energy effective theory. For
completeness, we recall the standard expressions for these masses.
For a general theory with
 $n$ chiral superfields, $Q^a$, with canonical classical K\"ahler potential,
$K_{cal}=Q_a^\dagger Q^a$, and superpotential $W(Q_a)$:
\eqn\mbmf{m_0^2=\pmatrix{W^{\dagger ac}W_{cb}&W^{\dagger abc}W_c\cr
W_{abc}W^{\dagger c}&W_{ac}W^{\dagger cb}},\qquad
m_{1/2}^2=\pmatrix{W^{\dagger ac}W_{cb}&0\cr 0&W_{ac}W^{\dagger
cb}},} with $W_c\equiv \partial W/\partial Q^c$, etc., and $m_0^2$
and $m_{1/2}^2$ are $2n\times 2n$ matrices.

\subsec{The basic O'Raifeartaigh model}

The basic model has three chiral superfields, $X$, $\phi _1$, and
$\phi _2$, with classical K\"ahler potential $K_{cl}=X^\dagger
X+\phi _1^\dagger \phi _1+\phi _2^\dagger \phi _2$,  and
superpotential
 \eqn\wor{W=\half hX \phi _1^2+h m\phi _1\phi _2 -h\mu ^2X.}
We denote the coefficient $f$ of the linear term as $f=-h\mu ^2$, with
$\mu$ having dimensions of mass, to make the mass dimension explicit,
and to simplify expressions.
This theory has a $U(1)_R$ symmetry, with $R(X)=2$, $R(\phi
_1)=0$, $R(\phi _2)=2$. The tree-level potential for the scalars
is, $V_{tree}=\left| F_X\right| ^2 + \left| F_{\phi _1} \right|
^2+ \left| F_{\phi _2}\right|^2$, with
 \eqn\for{ F_X=h\left(\half \phi _1^2 -\mu ^2\right), \quad F_{\phi
 _1}=h\left(
 X\phi _1 +m\phi _2\right), \quad F_{\phi_2}=h m\phi _1.}
Supersymmetry is broken because $F_X$ and $F_{\phi _2}$ cannot
both vanish. The $X$ and $\phi _2$ equations of motion require
that $F_{\phi _1}=0$, which fixes $\ev{\phi _2}=-\ev{X\phi _1/m}$.
The minimum of the potential is a moduli space of degenerate,
non-supersymmetric vacua, with $\ev{X}$ arbitrary. The minimum of
the potential depends on the parameter
 \eqn\ydef{y\equiv \left|{\mu ^2\over m^2}\right|}
For $y\leq 1$, the potential is minimized, with value
$V=|h^2\mu^4|$, at $\phi _1=\phi _2=0$ and arbitrary $X$. There is
a second order phase transition at $y=1$, where this minimum
splits to two minima and a saddle point.  For $y\ge 1$ the
potential has minima with $V=|h^2\mu^4|\left({2y-1\over
y^2}\right)$ at $\phi_1=\pm i \sqrt{{2\mu^2}(1-1/y)}$,
$\phi_2=-X\phi_1/m$ with arbitrary $X$. Let us focus on the $y\leq
1$ phase.

The fermion $\psi _X$ is the exactly massless Goldstino.  The
scalar component of $X$ is a classically pseudo-modulus.
The classical mass spectrum of the $\phi _1$ and $\phi _2$ field
can be computed from \mbmf.  For
the fermions, the eigenvalues are
 \eqn\eigenei{ m_{1/2}^2 = {1\over4} |h|^2(|X|\pm
 \sqrt{|X|^2+4|m|^2})^2,}
and for the real scalars the mass eigenvalues are
 \eqn\eigenii{ m_{0}^2 =|h|^2\left(|m|^2+\half \eta |\mu ^2|  +\half
 |X|^2\pm\half \sqrt{|\mu ^4|+2\eta |\mu ^2|
 |X|^2+4|m|^2|X|^2+|X|^4}\right), }
where $\eta =\pm 1$.  At $y=1$, where the second order phase
transition occurs, one of the eigenvalues \eigenii\ vanishes for all
$X$: the otherwise massive fields from $\phi _1$ and $\phi _2$ yield
an additional, classically massless, real scalar.

The classical flat direction of the classical pseudo-modulus $X$
is lifted by a quantum effective potential, $V_{eff}(X)$.  The
one-loop effective potential can be computed from the expression
\CWgen\ for the one-loop vacuum energy, using the classical masses
\eigenei\ and \eigenii. The pseudo-modulus $X$ is here treated as
a background. It is found that the resulting effective potential
is minimized at $\ev{X}=0$, so we'll simplify the expressions by
just expanding around this minimum:
$V_{eff}=V_0+m_X^2|X|^2+\dots$. The one loop corrected vacuum
energy is
 \eqn\voor{\eqalign{V_0&=|h^2\mu ^4|\left[1+{|h^2|\over
 64\pi^2}\left(y^{-2}(1+y)^2\log(1+y)+y^{-2}(1-y)^2\log(1-y)
 +2\log{|hm|^2\over\Lambda^2 }  \right)\right].}
 }
The dependence on the cutoff $\Lambda$ can be absorbed into the
running $h$. The one-loop quantum mass of the classical
pseudo-modulus $X$ is given by
 \eqn\oraighm{m_X^2=+{|h^4\mu ^2|\over 32\pi ^2}
 y^{-1}\left(-2+y^{-1}(1+y)^2\log (1+y)-y^{-1}(1-y)^2\log
 (1-y)\right). }
The mass \oraighm\ indeed satisfies $m_X^2>0$, consistent with the
minimum of the  one-loop potential \CWgen\ being at the origin.
For small supersymmetry breaking, $y\rightarrow 0$, we have
 \eqn\mxysmall{m_X^2\rightarrow {|h^4\mu ^4| \over 48\pi ^2
 |m|^2}, \qquad\hbox{for}\qquad |\mu ^2|\ll |m^2|.}
In the limit, $y\rightarrow 1$, where the supersymmetry breaking is large, we have
 \eqn\mxyi{m_X^2={|h^4\mu ^2|\over 16\pi ^2}(\log 4-1)
 \qquad\hbox{for}\qquad |\mu ^2|=|m|^2.}

Because the potential is minimized at $\ev{X}=0$, the vacuum has
broken supersymmetry but unbroken $U(1)_R$ symmetry.  If the
superpotential contains all terms allowed by symmetries, then
having a $U(1)_R$ symmetry is a necessary condition for
supersymmetry breaking, and having $U(1)_R$ spontaneously broken
is a sufficient condition for supersymmetry breaking \NelsonNF.
Here we find that the correct quantum vacuum is actually that
where $U(1)_R$ symmetry is not spontaneously broken, but
supersymmetry is nevertheless broken.

When the supersymmetry breaking mass splittings are small, the
effective potential can alternatively be computed in the
supersymmetric low-energy effective theory where we integrate out
the massive fields $\phi _1$ and $\phi _2$.  The effective
superpotential of the low-energy theory is $W_{low}=-h\mu ^2X$, and
the effective K\"ahler potential, $K_{eff}(X, X^\dagger)$, gets a
one-loop correction from integrating out the massive fields.  This
gives the effective potential
 \eqn\KeffV{V^{(1)}=(K_{eff\ XX^\dagger})^{-1}|h^2\mu ^4|.}
This way of computing the effective potential is valid only when
the supersymmetry breaking is small, because the true effective
potential generally gets significant additional contributions
{}from terms that involve higher super-derivatives in superspace.
The effective potential \CWgen\ gives the full answer, whether or
not the supersymmetry breaking is small. In particular, \KeffV\
only reproduces the effective potential \CWgen\ to leading order
in the $y\rightarrow 0$ limit.  For example, \KeffV\ reproduces
the mass \mxysmall\ of the small supersymmetry breaking limit, but
not the mass \mxyi\ of the large supersymmetry breaking limit.  In
appendix A.5 we prove, for generalized theories of tree-level
supersymmetry breaking, that the potential \KeffV, computed from
the effective K\"ahler potential, always agrees with the order
$|f|^2$ truncation of the correct effective potential, computed
via \CWgen.

\subsec{Some closely related examples}

Consider a theory of $2n+1$ chiral superfields, $X$, and $A_i$,
$B_i$, with $i=1\dots n$, K\"ahler potential $K=X^\dagger X+\sum
_iA_i^\dagger A_i+B_i^\dagger B_i$, and superpotential
 \eqn\worn{W=fX+\sum _i \left(\half h_i XA_i^2+h_i m_i
 A_iB_i\right) .}
This is not quite the same as $n$ decoupled copies of the
O'Raifeartaigh model \wor, because the same chiral superfield $X$
participates in each of them.  Taking all $y_i\equiv |f/h_i
m_i^2|\leq 1$, the classical vacuum is at $\ev{A_i}=\ev{B_i}=0$,
with $\ev{X}$ arbitrary and $V_{tree}=|f|^2$.  The fermion $\psi
_X$ is exactly massless, and the scalar component of $X$ gets mass
starting at one-loop.  The one-loop effective potential is
computed from the vacuum energy \CWgen, using the classical mass
spectrum computed as a function of $\ev{X}$.  The classical masses
of $A_i$ and $B_i$ come from expanding $V_{tree}$ to quadratic
order in the $A_i$ and $B_i$ fields (the general formula is given
in \mbmf). For example, for the scalars, we have
 \eqn\scarexp{V_{tree}\supset \sum
 _i\left({\rm Re}(f^*h_i A_i^2)+ \left|h_i m_i A_i\right|
 ^2+\left| h_i m_i B_i+\half h_i A_i X\right|
 ^2\right)
 .}
These masses are the same as in the original O'Raifeartaigh model
\wor, for each flavor $i$; the fermion masses are likewise simply
a sum of those of the model \wor, for each flavor $i$.

For each flavor $i$, the mass-squared eigenvalues are thus as in
\eigenei\ and \eigenii, and the one-loop effective potential
\CWgen\ is a simply a sum over $i$ of that of the original model
\wor; so the minimum of the effective potential is again at
$\ev{X}=0$. In particular, the one-loop quantum mass of $X$ is
given (with $y_i\equiv  |f/h_i m_i^2|$) by
 \eqn\oraighmn{m_X^2=\sum _{i=1}^n{|h_i^3f|\over 32\pi
 ^2}y_i^{-1}\left(-2+y_i^{-1}(1+y_i)^2\log (1+y_i)-
 y_i^{-1}(1-y_i)^2\log (1-y_i)\right). }

As another example, consider a theory of $2N$ chiral superfields
$S_i$ and $V_i$, $i=1\dots N$, with
$K=S^{i\dagger}S_i+V^{i\dagger}V_i$ and superpotential
 \eqn\wsvo{W=mS_i V_i, \qquad \hbox{subject to}\qquad V_iV_i=
 \Lambda ^2. }
There is an $SO(N)\times U(1)_R$ global symmetry, with $R(S_i)=2$
and $R(V_i)=0$.   It is impossible for $F_{S_i}=mV_i$ to all
vanish, because of the constraint $V_iV_i=\Lambda ^2$, so
supersymmetry is broken.  The constraint also spontaneously breaks
the $SO(N)$ flavor symmetry to $SO(N-1)$, so there are $N-1$
massless Goldstone bosons.  Solving the constraint equation, we
can take $\vec V\equiv (\sqrt{\Lambda ^2-\vec \phi _1 \cdot \vec
\phi _1},\vec \phi _1)$, and also define $\vec S\equiv (X, \vec
\phi _2)$, where $\vec \phi _1$ and $\vec \phi _2$ are $N-1$
component vectors. Writing the superpotential \wsvo\ to cubic
order, we have
 \eqn\wsvoo{W=m\Lambda X -\half {m\over \Lambda}
 X\vec \phi _1^2+m\vec \phi _1\cdot \vec \phi _2+\dots.}

The theory \wsvo\ now coincides with \worn, with $n=N-1$,
$m_i=\Lambda$, $h_i=-m/\Lambda$, and $f=m\Lambda$.  Because all
$y_i=|h_if/m_i^2|=1$, each component of the O'Raifeartaigh field
$\vec \phi _1$ includes a real massless scalar. In the present
model we identify them with the $SO(N)/SO(N-1)$ Goldstone bosons.
The one-loop mass \oraighmn\ is here
 \eqn\oraghmni{m_X^2=(N-1){|m|^4\over 16\pi ^2|\Lambda |^2}
 \left(\log 4 -1\right).}
For $N=6$, \wsvo\ is the effective macroscopic theory of the $SU(2)$ model,
with $N_f=2$ and $W=\lambda S^{ij}V_{ij}$, of  \refs{\IT, \IY}.
There $m=\lambda \Lambda$, with $\Lambda$ the dynamical scale of the $SU(2)$ gauge
theory, which also enters in the constraint \wsvo\ \SeibergBZ. For
this theory, essentially the above perturbative analysis, showing
that the one loop potential of the effective theory pushes the
pseudo-modulus to the origin, was given in \Chacko.

\subsec{Further generalizations}

More generally, let us couple a field $X$ to $N$ fields $\phi _i$
via:
 \eqn\wxh{W=fX+\half \phi _iM(X)^{ij}\phi _j.}
The example \wor\ has $M(X)=h\pmatrix{X&m\cr m&0}$, linear in $X$,
but more generally $M(X)$ need not be linear in $X$. Taking all
fields to have canonical K\"ahler potential, the classical
potential for the scalars is $V_{tree}=\left| F_X\right| ^2+
F_{\phi_i} ^\dagger F_{\phi _i}$, with
 \eqn\vxhcl{F_X=f+\half
 \phi _iM'(X)^{ij}\phi _j, \qquad F_{\phi _i} = M(X)^{ij} \phi _j.}

If $\det M(X)$ depends on $X$, then there will necessarily be
values $X=X_0$ where it vanishes, and then $F_{\phi _i}=0$ has a
solution for non-vanishing $\phi _i^0$.  In this case, there are
generally supersymmetric vacua.  These supersymmetric vacua could
be endpoints of runaway directions.   As a simple example with a
runaway, consider $W=fX+\half X^2\phi ^2$, with $F_X=f^2+X\phi ^2$
and $F_\phi =X^2\phi$. The potential has a runaway, to a
supersymmetric vacuum at $X=-f/\phi ^2$, with $\phi \rightarrow
0$.

If $\det M(X)$ is a non-zero, $X$ independent constant (as in the
model \wor), then the only solution of $F_{\phi _i} =0$ is $\phi
_i=0$.  If $\det M(X)$ is a non-zero constant, but $M(X)^{ij}$ is
not linear in $X$, then there is a possible runaway to a
supersymmetric vacuum; one must check the particular model in more
detail.  If $\det M(X)$ is a non-zero constant, and $M(X)^{ij}$ is
linear in $X$, then there is no runaway direction and
supersymmetry is broken, generalizing the O'Raifeartaigh model,
where $M(X)=h\pmatrix{X&m\cr m&0}$. If $\det M(X)$ vanishes
identically, then one must check further the particular model to
determine whether or not supersymmetry is broken.

\subsec{Comments about integrating out}

Consider a theory with $N$ chiral superfields $\phi_i$ and a
superpotential
 \eqn\wxh{W=\half \phi _iM^{ij}\phi _j +\quad\hbox{terms
 involving other fields}.}
We take $M^{ij}$ to be a symmetric matrix of background
superfields.  The other fields can lead to $M$ having a non-zero,
supersymmetry breaking, $F$ component, $F_M$.

We now integrate out $\phi_i$. The result is a supersymmetric
effective action for the background superfields $M^{ij}$. Because
our theory is quadratic in $\phi_i$, the effective K\"ahler
potential for $M^{ij}$ is exact at one-loop:
 \eqn\koneg{\eqalign{K_{eff}&=-{1\over 32\pi^2}\Tr
 \left[M^\dagger M\log (M^\dagger M/\Lambda ^2)\right]
 = -{1\over2}\Tr \int {d^4p\over (2\pi)^4}{1\over
 p^2+M^\dagger M} + const.}}
Here the integrals are regulated in the UV by $\Lambda$ and the
constant is proportional to $\Lambda^2$.  This expression is
familiar from the study of a theory with dynamical $M$, where it
arises from the one loop renormalization of the kinetic term of
$M$.

One way to see that \koneg\ is correct is to expand it in
components and focus on the term proportional to $F_MF_M^\dagger$:
 \eqn\konegintder{\eqalign{
 \int &d^4\theta K_{eff}\Big|_{F_MF_M^\dagger}  \cr
 &= -{1\over2}\Tr\int {d^4p\over
 (2\pi)^4}\left( \Delta^{-2}M^\dagger F_M \Delta^{-1}F_M^\dagger M
 + \Delta^{-2}F_M^\dagger M\Delta^{-1}M^\dagger F_M
 -\Delta^{-2}F_M^\dagger F_M\right) \cr
  &= -{1\over2}\Tr\int {d^4p\over (2\pi)^4}\left(
 \tilde \Delta^{-2}(\tilde \Delta-p^2) F_M \Delta^{-1}F_M^\dagger
 +  \Delta^{-2}F_M^\dagger \tilde\Delta^{-1}(\tilde\Delta-p^2) F_M
 -\Delta^{-2}F_M^\dagger F_M\right)\cr
 &= -{1\over2}\Tr\int {d^4p\over (2\pi)^4}\left(
 \tilde\Delta^{-1}F_M \Delta^{-1}F_M^\dagger +p^2{d\over dp^2}
 \Delta^{-1}F_M^\dagger \tilde\Delta^{-1} F_M\right)\cr
 &= +{1\over2}\Tr\int {d^4p\over (2\pi)^4}
 \tilde\Delta^{-1}F_M \Delta^{-1}F_M^\dagger
 }}
where $\Delta=p^2+M^\dagger M$ and $\tilde \Delta=p^2+MM^\dagger
$.  In the second line we used the fact that $Mf(M^\dagger M)
M^\dagger=f(MM^\dagger) MM^\dagger= MM^\dagger f(MM^\dagger)$ for
every function $f$, and in the last line we have integrated by
parts. The final result agrees with a one loop diagram with two
external fields $F_M$ and $F_M^\dagger$, and thus  confirms our
expression for  \koneg.

The full effective action includes terms which are higher order in
$F_M$ and $F_M^\dagger$.  Again, since the $\phi_i$ are free, they
can be integrated out exactly at one-loop, and then the full
effective action can be evaluated as a supertrace over the masses
of the particles,
 \eqn\colemanweinberg{{\cal L}_{eff}=-{1\over 64\pi ^2}{\rm Str}\ {\cal M}^4\log{
 {\cal M}^2\over \Lambda ^2}=-{1\over 4}{\rm Str}\int {d^4p\over (2\pi)^2}{{\cal M}^2\over p^2+{\cal M}^2}.}
The bosonic mass-squared matrix for the fields $\pmatrix{\phi &
\phi ^*}$ is $m_B^2=E+H$ with 
 \eqn\mmbex{E\equiv \pmatrix{M^\dagger M&0\cr 0&MM^\dagger}, \qquad H\equiv \pmatrix{0&F_{M}^\dagger\cr F_M&0},}
where the components are for $\pmatrix{\phi & \phi ^*}$ and
$\pmatrix{\phi ^* &\phi }^T$.  The fermion mass-squared matrix is
$m_F^2=E$.  If we expand \colemanweinberg\ in powers of $F_M$ and
$F_M^\dagger$, the leading term coincides with that obtained from
the effective K\"ahler potential \konegintder; to show this we
define $\Gamma \equiv p^2+E$,
 \eqn\cwexp{\eqalign{{\cal
L}_{eff}
 &=-{1\over 4}\Tr \int {d^4p\over (2\pi)^2}{m_B^2\over
p^2+m_B^2}+ {1\over 4}\Tr \int {d^4p\over (2\pi)^2}{m_F^2\over
p^2+m_F^2}\cr
 &=-{1\over 4}\Tr \int {d^4p\over
(2\pi)^2}\left[(E+H)(1+\Gamma ^{-1}H)^{-1}-E\right]\Gamma ^{-1}\cr
 &=-{1\over 8}\Tr \int {d^4p\over
(2\pi)^2}p^2{d\over dp^2}\left(H\Gamma ^{-1}\right)^2 +{\cal
O}(F^4),\cr
 &= +{1\over2}\Tr\int {d^4p\over (2\pi)^4}
 \tilde\Delta^{-1}F_M \Delta^{-1}F_M^\dagger+{\cal O}(F^4).
 }}
This agrees with the expression \konegintder, coming from the
effective K\"ahler potential \koneg.  However, \koneg\ does not
capture the terms of higher order in $F_M$ and $F_M^\dagger$ in
the first two lines of \cwexp.

\appendix{B}{Calculating $a$ and $b$}

\subsec{$SU(N)$ case}

In this appendix, we flesh out the calculation of the one-loop
effective potential \Vonegen\ on the pseudo-moduli space of the
$SU(N)$ macroscopic theory.  As noted in section 2, this
calculation reduces to determining two numerical coefficients $a$
and $b$. More generally, the one-loop potential is computed from
the one-loop vacuum energy \CWgen, treating the pseudo-moduli as a
classical background. It thus suffices to expand away from the
vacuum \maximal\ along a two parameter space labelled by $X_0$ and
$\theta$: \eqn\mqfluctsunge{ \Phi =\pmatrix{ \delta Y& \delta
Z^T\cr \delta\tilde Z& X_0\unit_{N_f-N}+\delta \hat\Phi}, \quad
\varphi=\pmatrix{\mu e^{\theta}\unit_N+\delta\chi\cr
\delta\rho},\quad \tilde \varphi ^T=\pmatrix{ \mu
e^{-\theta}\unit_N+\delta\tilde\chi\cr \delta\tilde \rho},
 }
with $X_0$ and $\theta$ treated as small parameters. To compute
\CWgen, we need the classical masses of the fluctuations in
\mqfluctsunge, as functions of the small pseudo-moduli background.
This yields the one-loop correction to the vacuum energy,
\eqn\vaccorrsun{ \left\langle V_{eff}^{(1)}\right\rangle = const.
+ h^4\mu^2 \left({1\over2}a\,N \mu^2(\theta+\theta^*)^2 +
b(N_f-N)|X_0|^2\right) + \dots,
 }
from which we can read off the coefficients $a$ and $b$. (For
simplicity we take $h$ and $\mu$ real and positive throughout this
appendix.)

To compute the classical masses, we substitute
\mqfluctsunge\ into the superpotential \Wsunge:
\eqn\Wexpand{\eqalign{
W &= h\Tr\, \varphi \Phi \tilde \varphi - h\, \mu^2\Tr\,\Phi \cr
 &= h\Tr\,\Bigg[
 \mu e^{\theta}\delta Z^T\delta\tilde\rho+\mu e^{-\theta}\delta\tilde
   Z^T\delta\rho
  + \delta\rho^T(X_0+\delta \hat\Phi)\delta\tilde\rho-\mu^2(X_0+\delta
  \hat\Phi)\cr
   &\qquad\qquad +\mu e^{\theta}\delta Y\delta\tilde\chi+\mu e^{-\theta}\delta Y^T\delta\chi
 \Bigg]+ \dots
 }}
where $\dots$ contains terms of cubic order and higher in the
fluctuations. According to \Wexpand, the off-diagonal components
of $\delta\hat\Phi$ do not contribute to the mass matrix, so we can
neglect them here.  Moreover, the fields
$\delta\chi$, $\delta\tilde\chi$, and $\delta Y$ only couple to
the supersymmetry breaking fields $\delta\rho$ and
$\delta\tilde\rho$ through terms of cubic or higher order in the
fluctuations. Therefore, the mass matrix for these fields will be
supersymmetric, and they will not contribute to the supertrace. So
they can also be neglected here. The remaining relevant terms are
\eqn\Wexpandred{
W\supset h\sum_{f=1}^{N_f-N}\left[ (X_0+\delta
\hat\Phi_{ff})(\delta\rho\delta\tilde\rho^T)_{ff}
 +\mu e^{\theta}(\delta\tilde\rho\delta Z^T)_{ff}+\mu e^{-\theta}(\delta\rho\delta\tilde Z^T)_{ff}-\mu^2
 (X_0+\delta \hat\Phi_{ff})\right].
}
We recognize $N_f-N$ decoupled copies of an O'Raifeartaigh-like
model of the form
\eqn\ORsun{
W = h \left(X \vec\phi_1\cdot\vec\phi_2 +\mu
e^{-\theta}\vec\phi_1\cdot\vec\phi_3+\mu
e^{\theta}\vec\phi_2\cdot\vec\phi_4 -\mu^2 X\right)
}
where the $\vec\phi_i$ are $N$ dimensional vectors. A calculation
completely analogous to those in appendix A yields the one-loop
vacuum energy coming from these $N_f-N$ O'Raifeartaigh-like
models, as a function of $\langle X\rangle = X_0$ and $\theta$. We
find:
\eqn\Veffsunspec{
\left\langle V^{(1)}_{eff}\right\rangle = const.
 + {h^4\mu^2(\log 4-1)N(N_f-N)\over 8\pi^2}\left(
 {1\over2}\mu^2(\theta+\theta^*)^2+|X|^2 \right)
  + \dots
}
Comparing with \vaccorrsun, we read off the coefficients $a$ and
$b$:
\eqn\aandbspec{
a = {\log 4-1\over 8\pi^2}(N_f-N),\qquad b = {\log 4-1\over
8\pi^2}N
}
This is the answer \abans\ quoted in section 2.

\subsec{$SO(N)$ case}

The $SO(N)$ macroscopic model studied in section 6 can also be
analyzed along the lines of the previous subsection. To begin, we
expand around a point near \maximalson,
\eqn\nearbypointson{
\Phi_0 = X_0\unit_{N_f-N},\qquad \varphi_0=\mu
\pmatrix{\cosh\theta & i\sinh\theta\cr -i\sinh\theta &
\cosh\theta}\otimes \unit_{N/2}
 }
where for simplicity we are assuming $N$ is even. The general form
of the one-loop vacuum energy, expanded around $X_0=\theta=0$, is
\eqn\vaccorrson{
\left\langle V_{eff}^{(1)}\right\rangle =
 const. +
h^4\mu^2\left({1\over8}a\,N \mu^2(\theta+\theta^*)^2 +
b(N_f-N)|X_0|^2\right) + \dots
 }
To calculate the coefficients $a$ and $b$, we reduce the
superpotential as in the previous subsection, yielding the relevant terms
\eqn\Wsonred{\eqalign{
W&\supset
  h\sum_{f=1}^{N_f-N}\Bigg[
  (X_0+\delta\hat\Phi_{ff})(\delta\rho \delta\rho^T)_{ff}+\sqrt{2}(\delta\rho\varphi_0^T\delta Z^T)_{ff}
  -\mu^2(X_0+\delta\hat\Phi_{ff})\Bigg]
 }}
This is equivalent to $N_f-N$ decoupled copies of the
O'Raifeartaigh-like model
\eqn\ORson{
W = h\left[X(\vec\phi_1^2+\vec\phi_2^2) +\sqrt{2}\mu
\pmatrix{\vec\phi_1\cr\vec\phi_2}^T\pmatrix{\cosh\theta &
-i\sinh\theta\cr i\sinh\theta &
\cosh\theta}\pmatrix{\vec\phi_3\cr\vec\phi_4}-\mu^2 X\right]
}
where the $\vec\phi_i$ are $N/2$ dimensional vectors. By a unitary
transformation,
\eqn\unitaryson{
(\vec\phi_1,\vec\phi_2,\vec\phi_3,\vec\phi_4)\to
\left(-{i(\vec\phi_1-\vec\phi_2)\over\sqrt{2}},\,
{\vec\phi_1+\vec\phi_2\over\sqrt{2}},\,
 {i(\vec\phi_3-\vec\phi_4)\over\sqrt{2}},\,
  {\vec\phi_3+\vec\phi_4\over\sqrt{2}}\right)
  }
we can actually turn \ORson\ into
\eqn\ORsonii{
W = h\left( 2X\vec\phi_1\cdot\vec\phi_2 +\sqrt{2}\mu
e^{-\theta}\vec\phi_1\cdot\vec\phi_3+\sqrt{2}\mu
e^{\theta}\vec\phi_2\cdot\vec\phi_4 - \mu^2 X
 \right)
 }
which is the O'Raifeartaigh-like model of the previous subsection,
but with $\mu_{here}=\sqrt{2}\mu_{there}$ and
$h_{here}={1\over2}h_{there}$. Therefore, we can copy over the
vacuum energy from the previous subsection, rescaled
appropriately:
\eqn\Veffsonspec{
\left\langle V^{(1)}_{eff}\right\rangle = const.
 + {h^4\mu^2(\log 4-1)N(N_f-N)\over 2\pi^2}\left(
 {1\over4}\mu^2(\theta+\theta^*)^2+|X|^2 \right)
  + \dots
}
Comparing with \vaccorrson, we can read off $a$ and $b$. The
result is the answer \Vonegensonii\ quoted in section 6.

\subsec{$Sp(N)$ case}

Finally, let us analyze the $Sp(N)$ macroscopic model of section 6
in the same way. We expand around a point near \maximalspn,
\eqn\nearbypointspn{ \Phi_0 = X_0\unit_{N_f-N}\otimes
(i\sigma_2),\qquad \varphi_0=\mu \pmatrix{\cosh\theta &
i\sinh\theta\cr -i\sinh\theta & \cosh\theta}\otimes \unit_N }
(Recall our conventions are such that $J_{2N}=\unit_N\otimes
(i\sigma_2)$.) The general form of the one-loop vacuum energy,
expanded around $X_0=\theta=0$, is \eqn\vaccorrspn{ \left\langle
V_{eff}^{(1)}\right\rangle = h^4\mu^2\left({1\over2}2N
a\mu^2(\theta+\theta^*)^2+2(N_f-N)b|X|^2\right)+\dots } To
calculate \vaccorrspn, we again expand the superpotential and
reduce it as in the previous subsections. This yields precisely
the same O'Raifeartaigh model \ORson\ as for $SO(N)$, except with
$(N,N_f-N)$ in $SO(N)$ replaced with $(4N,N_f-N)$. Therefore, the
one-loop vacuum energy is just \Veffsonspec\ multiplied by four,
\eqn\Veffspnspec{ \left\langle V_{eff}^{(1)}\right\rangle = const.
 + {2h^4\mu^2(\log 4-1)N(N_f-N)\over \pi^2}\left(
 {1\over4}\mu^2(\theta+\theta^*)^2+|X|^2 \right)
  + \dots
  }
Comparing with the general form \vaccorrspn\ and reading off $a$
and $b$, we obtain the answer \Vonespnii\ quoted in the text.

\appendix{C}{A landscape of supersymmetry breaking vacua}

Consider $\CN =1$ supersymmetric SQCD, with gauge group $SU(N_c)$
and $N_f$ flavors, and add an extra chiral superfield $\Phi$ in
the adjoint representation, with superpotential (see e.g.\
\refs{\DKi\DKAS-\DKNSAS})
 \eqn\wdeltanix{W=\sum _{p=1}^{K+1} {1\over p}\Tr g_p\Phi ^p+ \Tr m M.}
(For simplicity we do not include superpotential terms coupling
$\Phi$ to the fundamentals.  They can be easily added.) Let us
consider the case of large $g_p$, where we should expand around
the classical vacua of \wdeltanix. There is a ``landscape" of such
classical vacua, with $SU(N_c)$ Higgsed by the $\ev{\Phi}$ as
 \eqn\breaking{U(N_c)\rightarrow \prod _{i=1}^K U(N_i)
 \qquad\hbox{for all partitions}\qquad N_c=\sum _{i=1}^K N_i; \qquad
 N_i\geq 0.}
The number of such possibilities grows rapidly with $K$ and $N_c$.

For generic and large $g_p$, all of the components of $\Phi$ in
each of these vacua are massive. The low-energy theory in each
vacuum consists of approximately decoupled $U(N_i)$ gauge groups.
Each $U(N_i)$ group has $N_f$ flavors, with identical masses given
by $m$ in \wdeltanix. Suppose now that at least one $N_i$
satisfies
 \eqn\Nireq{ N_i+1\le N_f< {3\over 2}N_i }
then, using the analysis in sections 2 -- 5, the $U(N_i)$ theory
has meta-stable supersymmetry breaking vacua.  We see that this
theory has many supersymmetric as well as many compact spaces of
meta-stable vacua. There is thus a landscape of supersymmetric and
meta-stable non-supersymmetric vacua.

Such vacua are also present in the string theory landscape, as
these gauge theories have string realizations.  In this context
the integers $N_i$ arise as the number of branes or the values of
certain fluxes.

As an aside, we note that one can also construct field theory
examples with a landscape of non-supersymmetric vacua, with no
supersymmetric vacuum.  Consider, for example the supersymmetry
breaking model of \AffleckXZ, based on $SU(N_c)$ gauge theory,
with $N_c$ odd, and matter in the $\doub \oplus (N_c-4)
\overline{\square}$.  As noted in  \refs{\IntriligatorFK,
\LeighSJ}, it is interesting to consider adding to this theory an
adjoint $\Phi$, with superpotential as in \wdeltanix. We again get
a classical landscape of vacua for $\Phi$, with the breaking
patterns \breaking.  In some of these vacua, the low-energy theory
reduces to one that was already known to break supersymmetry
\refs{\IntriligatorFK, \LeighSJ}.   A priori, one might expect
that some of the vacua break supersymmetry, and others might not.
A systematic analysis has not yet been completed, but it seems
possible that every vacuum of the classical landscape of
\breaking\ breaks supersymmetry in this present case.

\appendix{D}{${\cal N}=2$ Super Yang-Mills, slightly broken to
${\cal N}=1$.}

In ${\cal N}=2$ supersymmetric gauge theory, the exact K\"ahler
potential of the low-energy effective theory on the Coulomb branch
can be determined, from a holomorphic quantity (the prepotential)
\SeibergRS. Let us consider an $\CN =2$ theory, broken to $\CN =1$
by superpotential terms,
 \eqn\wdeltanii{\Delta W_{tree}=\sum _p {1\over p}g_p\Tr \Phi ^p\equiv
 \sum _p g_p u_p.}
The supersymmetric vacua of this theory have been much studied
(see e.g.\ \refs{\CachazoJY\DijkgraafDH- \CachazoRY}). We can also
look for meta-stable minima of the effective potential on the
Coulomb branch, \eqn\veff{V_{eff}=\sum _{p\overline
p}(K^{-1}_{eff})^{u_pu^\dagger _{\overline p}}g_pg^*_{\overline
p}.} Taking all $g_p\ll 1$, where $\CN =2$ is just slightly broken
to $\CN =1$, we can use the exactly determined $\CN =2$ K\"ahler
potential $K_{eff}(u_p, u_p^\dagger, \Lambda)$ in \veff, to get
the effective potential to leading order in $g_p$, but exactly in
$\Lambda$.    We can there look for meta-stable vacua, without the
ambiguity of the order one coefficients $\alpha$ and $\beta$ that
appeared in section 5.

For example, consider $\CN =2$ supersymmetric $SU(2)$ Yang-Mills
theory, broken to $\CN =1$ as in \wdeltanii\ by a mass term
$g_2=m_\Phi$.  For $g_2=0$,  the low-energy effective theory is an
${\cal N}=2$ $U(1)$ vector multiplet.  There is a moduli space of
${\cal N}=2$ supersymmetric vacua, with K\"ahler metric given by
\SeibergRS\
 \eqn\kmetis{ds^2={\rm Im}\tau |da|^2, \qquad
 \tau = {da_D/du\over da/du},} with \eqn\aadis{a(u)={\sqrt{2}\over
 \pi}\int _{-1}^1{dx\sqrt{x-u}\over \sqrt{x^2-1}},
 \quad\hbox{and}\quad a_D={\sqrt{2}\over \pi} \int _1^u
 {dx\sqrt{x-u}\over \sqrt{x^2-1}}.}
The functions $a(u)$ and $a_D(u)$ can be expressed in terms of
hypergeometric functions. The dynamical scale $\Lambda$ was set to
unity; it can be restored by dimensional analysis. Adding
$W_{tree}=m_\Phi u$ leads to supersymmetric vacua at $u=\pm 1$,
where a massless monopole or dyon condenses \SeibergRS.   We here
ask if there could also be meta-stable, non-supersymmetric vacua,
at other values of $u$.  In this case, it turns out that the
answer is no.

For small $m_\Phi $, the scalar potential is
 \eqn\vniiis{V_{eff}(u)=\left({\rm Im}\, \tau (u)\right)^{-1}
 \left|{da\over du}\right|^{-2} |m_\Phi|^2.}
it is straightforward to find that the only minima are the global
ones, at $u=\pm 1$.  There is a saddle point at $u=0$, where the
potential curves up along the ${\rm Im}\,u$ axis, but down along
the ${\rm Re}\, u$ axis.  The vacuum at $u=0$ is unstable to
rolling along the ${\rm Re}\, u$ axis, down to the minima at
$u=\pm 1$.

More generally, one could look for meta-stable non-supersymmetric
vacua in ${\cal N}=2$ supersymmetric $SU(N_c)$ SQCD, with $N_f$
massive flavors, slightly broken to ${\cal N}=1$ by \wdeltanii.
For $g_p=0$, the effective theory of the Coulomb branch, and in
particular the K\"ahler potential, are exactly given by the curve
$y^2=\det (x-\Phi)^2-\Lambda ^{2N_c-N_f}\prod _{f=1}^{N_f}(x+m_i)$
\refs{\tyui\tyuii\tyuiii\tyuiv-\tyuv}, where $m_i$ are the masses
of the flavors. Taking $g_2=m_\Phi$ to infinity, the low-energy
theory at $\Phi =0$ is governed by $\CN =1$ SQCD. There, as we
have argued, there are meta-stable, supersymmetry breaking vacua
for $N_f<{3\over 2}N_c$. Perhaps the meta-stable vacua can also be
seen in the opposite limit, where the $\CN =2$ breaking terms
\wdeltanii\ are small, and the infrared theory can be
approximately described using the exactly known $\CN =2$
supersymmetric K\"ahler potential.

\listrefs

\end